\documentclass[pra,superscriptaddress,reprint,amsfonts,amssymb,showpacs]{revtex4-1}
\usepackage{graphicx}
\usepackage{bm}
\usepackage{ulem}
\usepackage{color}
\usepackage{amssymb}
\usepackage{color}
\usepackage{hyperref}
\renewcommand{\sout}[1]{}
\renewcommand{\xout}[1]{}
\newcommand{\jap}[1]{}

\definecolor{WildStrawberry}{cmyk}{0,0.96,0.39,0}
\definecolor{Orange}{cmyk}{0,0.61,0.87,0}
\definecolor{Mahogany}{cmyk}{0,0.85,0.87,0.35}

\begin{document}

\title{Quantum magnonics: magnon meets superconducting qubit}

\author{Yutaka Tabuchi}
\email{tabuchi@qc.rcast.u-tokyo.ac.jp}
\author{Seiichiro Ishino}
\author{Atsushi Noguchi}
\author{Toyofumi Ishikawa}
\author{Rekishu Yamazaki}
\author{Koji Usami}
\affiliation{Research Center for Advanced Science and Technology (RCAST), The University of Tokyo, Meguro-ku, Tokyo 153-8904, Japan}
\author{Yasunobu Nakamura}
\affiliation{Research Center for Advanced Science and Technology (RCAST), The University of Tokyo, Meguro-ku, Tokyo 153-8904, Japan}
\affiliation{Center for Emergent Matter Science (CEMS), RIKEN, Wako, Saitama 351-0198, Japan}

\date{\today}


\begin{abstract}
The techniques of microwave quantum optics are applied to collective spin excitations in a macroscopic sphere of ferromagnetic insulator. 
We demonstrate, in the single-magnon limit, strong coupling between a magnetostatic mode in the sphere and a microwave cavity mode.
Moreover, we introduce a superconducting qubit in the cavity and couple the qubit with the magnon excitation via the virtual photon excitation.
We observe the magnon-vacuum-induced Rabi splitting. The hybrid quantum system enables generation and characterization of non-classical quantum states of magnons.     
\end{abstract}

{\maketitle}

\section{Introduction}
\label{intro}

The successful development of superconducting qubits and related circuits has brought wide opportunities in quantum control and measurement in the microwave domain~\cite{bib:Schoelkopf08,bib:You11,bib:Devoret13,bib:Riste13,bib:Weber14,bib:Kelly15}. In circuit quantum electrodynamics and microwave quantum optics, bosonic excitations of the electromagnetic modes, i.e., ``photons'' are handled with high accuracy~\cite{bib:Hofheinz09,bib:Flurin12,bib:Lang13,bib:Leghtas15}\footnote{To be more precise, we may say that surface plasmon polaritons, i.e., quanta of the hybridized modes of the surface charge density waves on the electrodes and the electromagnetic waves in the vacuum, are manipulated in the circuits.}. Therefore, it is natural to extend the targets to other quantum mechanical degrees of freedom. The examples are found in recent reports on hybrid quantum systems based on superconducting circuits: For example, paramagnetic spin ensembles~\cite{bib:Zhu11,bib:Kubo11}, nanomechanical oscillators~\cite{bib:Connell10,bib:Lecocq15,bib:Pirkkalainen15}, and surface acoustic waves in a piezoelectric substrate~\cite{bib:Gustafsson14}, have been coherently controlled via a coupling with a superconducting qubit.

Our goal here is to apply the techniques of microwave quantum optics to collective spin excitations in ferromagnet. Similar to superconductivity, ferromagnetism has a rigidity in its order parameter. The lowest energy excitations are long-wavelength collective spin precessions. We couple the quantum of the collective mode, a magnon, to a microwave cavity as well as a superconducting qubit to reveal its coherent properties in the quantum limit~\cite{bib:Tabuchi14,bib:Tabuchi15}. 

This paper is structured as follows: Section~2 reviews the basics of magnons in ferromagnet. In Sec.~3, hybridization of a magnon and a photon in a microwave cavity is demonstrated. Finally, in Sec.~4, we demonstrate strong coupling between a superconducting qubit and a magnetostatic mode in a ferromagnetic crystal. The magnon vacuum induces Rabi splitting in the qubit excitation.
Summary and outlook are presented in Sec.~5.

\section{Magnons in ferromagnet}
\label{theory}
\subsection{Spin waves} \label{sec:th_spin_wave}

In order to describe spin waves, or collective excitations in ferromagnetic materials, we begin with a simple Hamiltonian:
\begin{equation}
  \hat{{\cal H}}=
    -g\mu_{\mathrm{B}} B_z \sum_i \hat{S}_i^z 
    -2J\sum_{\langle i,j\rangle}\hat{\bf{S}}_i\cdot \hat{\bf{S}}_j,
  \label{eq:th_h-b}
\end{equation}
where the first term represents the Zeeman energy and the second one is the nearest-neighbor exchange interaction. 
The sum in the second term is taken over the pairs of the neighboring spins. $\hat{\mathbf{S}}_i$ is the Heisenberg spin operator for the $i$-th site, $g$ is the $g$-factor, $\mu_{\mathrm{B}}$ is the Bohr magneton, $B_z$ is the static magnetic field along the $z$ axis, and $J$ is the exchange integral. 
$J$ takes positive values for ferromagnetic materials,  leading to the ferromagnetic ground state, where all the spins are aligned along the $z$ axis.

\begin{figure}[htb]
  \centering
  \includegraphics{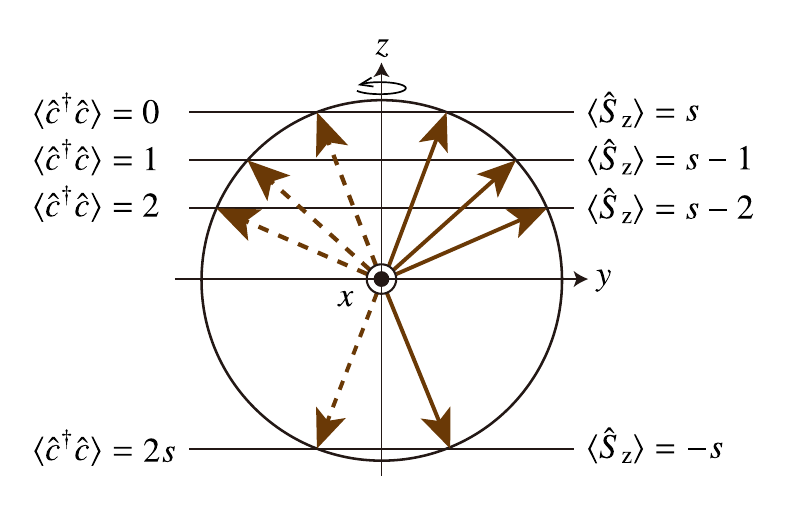}
  \caption{\label{fig:th_hpt}Relation between magnon number and $z$-component of total spin. The number of the bosons corresponds to the reduction of the total spin.}
\end{figure}

We can express the Heisenberg operators in terms of the bosonic operators $\hat{c}_i,\,\hat{c}_i^\dag$ by using the Holstein-Primakoff transformation~\cite{bib:Holstein40}:
\begin{eqnarray}
  \hat{S}_i^+&=& \hat{S}_i^x+i \hat{S}_i^y
  =
  \sqrt{2s} \left(
              1-\frac{\hat{c}_i^\dag \hat{c}_i}{2s}
            \right)^{1/2} \hat{c}_i, \label{eq:th_sip} \\
  \hat{S}_i^- &=& \hat{S}_i^x-i\hat{S}_i^y
  =
  \sqrt{2s}\, \hat{c}_i^\dag 
    \left(
      1-\frac{\hat{c}_i^\dag \hat{c}_i}{2s}
    \right)^{1/2}, \label{eq:th_sim} \\
  \hat{S}_i^z &=& s-\hat{c}_i^\dag \hat{c}_i, \label{eq:th_siz}
\end{eqnarray}
where $s$ is the total spin on each site. The meaning of this transformation is illustrated in Fig.~\ref{fig:th_hpt}. We find from Eq.~(\ref{eq:th_siz}) that the number of the bosons corresponds to the reduction of the $z$-component of the total spin.

The bosonic operators defined on each lattice point are related to the spin-wave operators by the Fourier transformation:
\begin{eqnarray}
	\hat{c}_i&=&\frac{1}{\sqrt{N}}\sum_{\bf{k}}e^{-i{\bf{k}}\cdot{\bf{r}}_i}\hat{c}_{\bf{k}}, \\
        \hat{c}_i^\dag&=&\frac{1}{\sqrt{N}}\sum_{\bf{k}}e^{i{\bf{k}}\cdot{\bf{r}}_i}\hat{c}_{\bf{k}}^\dag, 
\end{eqnarray}
where $N$ is the number of the atoms with spin $s$, and $\hat{c}_{\bf{k}}$ and $\hat{c}_{\bf{k}}^\dagger$ correspond to annihilation and creation of a magnon in the plain-wave mode, respectively. 
Substituting these operators into the Hamiltonian [Eq.~(\ref{eq:th_h-b})] and truncating it to the second order, we obtain the spin-wave Hamiltonian:
\begin{equation}
	\hat{\mathcal{H}}=\sum_{\bf{k}}\hbar\omega_{\bf{k}}\hat{c}_{\bf{k}}^\dag\hat{c}_{\bf{k}},
\end{equation}
with the dispersion relation:
\begin{equation}
  \hbar\omega_{\bf{k}}=2sZJ (1-\gamma_{\bf{k}})+g\mu_{\mathrm{B}} B_z,
  \label{eq:th_dispersion}
\end{equation}
where $Z$ is the coordination number of each site. In the case of a simple cubic lattice ($Z=6$) with the lattice constant $a_0$, $\gamma_{\bf{k}}$ becomes
\begin{equation}
  \gamma_{\bf{k}}=\frac{1}{3} (\cos k_x a_0+\cos k_y a_0+\cos k_z a_0 )  ,
\end{equation}
which gives the quadratic dispersion relation in the long wavelength limit:
\begin{equation}
  \hbar\omega_{\bf{k}} = 2sJa_0^2 |{\bf{k}}|^2+g\mu_{\mathrm{B}} B_z  .
\end{equation}
As indicated in the first term, The rigidity of the ordered spin system lifts the degeneracy of the spin excitations, which is in stark contrast with the case in paramagnetic spin ensembles.

%
%

\subsection{Magnetostatic modes}

\label{sec:magneto}

\begin{figure}[htb]
  \centering
  \includegraphics{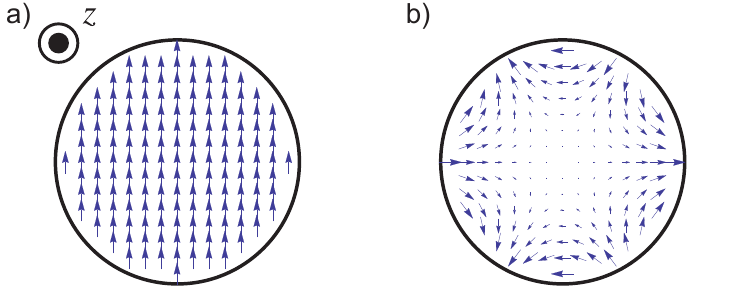}
  \caption{\label{fig:th_walker}Geometries of the transverse magnetization ${\bf{m}}$ in a) the Kittel $[(1,1,0)]$ mode and b) the $(4,3,0)$ mode. }
\end{figure}

While we assumed infinite lattices in Sec.~\ref{sec:th_spin_wave}, 
we actually need to consider finite samples coupling to the surrounding electromagnetic field. Especially when the wavelength of magnons is comparable to the sample size, the effect of the dipolar field generated by the spins becomes dominant, and thus the boundary conditions at the sample surface are of great importance. Relatively, the contribution of the exchange interactions ($\propto |\mathbf{k}|^2$) becomes negligible. We then determine the magnetization oscillation modes from classical electrodynamics. 

Suppose the magnetization in the sample is forced to oscillate by a time-dependent magnetic field ${\bf{h}}e^{i\omega t}$ perpendicular to the static magnetic field, where ${\bf{h}}=(h_x,h_y,0)$. Then the oscillating part ${\bf{m}} e^{i\omega t}=(m_x,m_y,0)e^{i\omega t}$ of the magnetization obeys the Landau-Lifshitz equation:
\begin{eqnarray}
  \frac{i\omega m_x}{\gamma} &=&m_y B_z-\mu_0M_\mathrm{s} h_y 
                                                   \label{eq:th_imx},\\
  \frac{i\omega m_y}{\gamma} &=&-m_x B_z+\mu_0M_\mathrm{s} h_x, 
                                                   \label{eq:th_imy} 
\end{eqnarray}
where $M_\mathrm{s}$ is the saturation magnetization, $\mu_0$ is the vacuum permeability, and $\gamma \equiv g \mu_\mathrm{B} / \hbar $ is the electron gyromagnetic ratio. 
Note that we have linearized these equations assuming that the amplitudes of $\bf{m}$ and $\bf{h}$ are small compared to $M_\mathrm{s}$ and $B_z/\mu_0$, respectively. 
Solving Eqs.~(\ref{eq:th_imx}) and (\ref{eq:th_imy}) for $m_x$ and $m_y$ yields the following results:
\begin{widetext}
\begin{eqnarray}
 m_x &=& 
   \frac{\mu_0M_\mathrm{s}}{B_z^2-(\omega/\gamma)^2 } 
    \left(B_z h_x-i \frac{\omega}{\gamma} h_y \right)
   \equiv 
      \kappa \frac{\partial\psi}{\partial x}
     -i\nu \frac{\partial\psi}{\partial y}, 
      \label{eq:th_mxm}\\
 m_y&=&
    \frac{\mu_0M_\mathrm{s}}{B_z^2-(\omega/\gamma)^2 } 
    \left(i \frac{\omega}{\gamma} h_x+B_z h_y \right) 
    \equiv 
       i\nu \frac{\partial\psi}{\partial x}
     + \kappa \frac{\partial\psi}{\partial y},
       \label{eq:th_mym} \\
  \kappa &=& 
    \frac{\mu_0M_\mathrm{s} B_z}{B_z^2-(\omega/\gamma)^2 },
      \quad \nu=\frac{\mu_0M_\mathrm{s} \omega/\gamma}{B_z^2-(\omega/\gamma)^2 },
   \nonumber
\end{eqnarray}
\end{widetext}
where $\psi$ is the scalar potential of $\bf{h}$ ($\nabla \psi \equiv \mu_0 \bf{h}$). Substituting Eqs.~(\ref{eq:th_mxm}), (\ref{eq:th_mym}) into the Maxwell equations:
\begin{equation}
	\nabla^2 \psi = \mu_0\,\mathrm{div}\, {\bf{h}}=-\mathrm{div}\, {\bf{m}},
\end{equation}
we obtain the differential equation for $\psi$ inside the sample:
\begin{equation}
  \left[
   (1+\kappa) 
    \left(
      \frac{\partial^2}{\partial x^2 }
     +\frac{\partial^2}{\partial y^2}
    \right)
   +\frac{\partial^2}{\partial z^2 }
  \right]\psi=0.
\end{equation}
Outside the sample, 
$\nabla ^2\psi = 0$. 
The boundary conditions are the continuities of $\psi$ and the normal component of ${\bf{h}} + {\bf{m}}/\mu_0$. Walker solved these equations for spheroidal samples and found for each eigenmode the mode shape and the $B_z$-dependence of the eigenfrequency~\cite{bib:Walker57,bib:Walker58,bib:Fletcher59}. These modes, bunched in a frequency range and characterized by three integer indices, are called magnetostatic or the Walker modes. The simplest is the $(1,1,0)$ mode, or the so-called Kittel mode, where all the spins in the sample precess in phase and with the same amplitude. The geometries of the transverse magnetization ${\bf{m}}$ in the Kittel mode and one of its degenerate modes $[(4,3,0)]$ 
are shown in Fig.~\ref{fig:th_walker}.

\subsection{Magnon linewidth}

Here we only consider magnons in insulating ferromagnets. 
Metallic ones suffer from strong damping of magnons due to scattering by the conduction electrons.

A number of magnon relaxation mechanisms are known in ferromagnetic insulators~\cite{bib:Sparks64,bib:Gurevich96}.
At high temperatures around room temperature, magnon-magnon and magnon-phonon inelastic scatterings are dominant because of the large number of thermally excited magnons and phonons.
The intrinsic magnon-magnon scattering is caused by the nonlinearity in the Holstein-Primakoff transformation [Eqs.~(\ref{eq:th_sip})-(\ref{eq:th_siz})], while the latter is caused by the spin-lattice coupling~\cite{bib:Kasuya61}. However, both mechanisms are negligible at low temperatures we are interested in.

At lower temperatures, extrinsic relaxation mechanisms become dominant. They are induced by the defects and impurities inside the crystals as well as the macroscopic pores and roughness at the surfaces. In the intermediate temperature range, typically between 10--100 K, the linewidth often shows a large peak in the temperature dependence. The peak height strongly depends on the amount of the defects such as rare-earth impurities and oxygen vacancies. It is known that the so-called slow-relaxation mechanism caused by the magnetic impurities is responsible for the broadening~\cite{bib:Teale62}. The effect of such mechanism also diminishes at lower temperatures.

A few relaxation mechanisms remain even at the zero-temperature limit. For instance, van Vleck's theory~\cite{bib:Vleck64} assumes an interaction of magnons with an ensemble of two-level systems (TLSs) which have the same excitation frequency as magnons. The relaxation rate is predicted to be proportional to $\tanh(\hbar\omega_{\mathrm{m}}/ 2 k_{\mathrm{B}}T)$, where $ k_{\mathrm{B}}$ is the Boltzmann constant. The characteristic temperature dependence is derived from the fact that the TLSs are saturated as the temperature increases. To the best of our knowledge, however, there has not been any observation of such temperature dependence in ferromagnetic resonance linewidth.

Another relaxation mechanism independent of the temperature is the elastic scattering of the Kittel-mode magnons due to the surface roughness of the samples~\cite{bib:Sparks61}. 
The surface roughness causes intermode coupling between the Kittel mode and other magnetostatic and spin-wave modes.

\section{Hybridization with a microwave cavity mode}\label{cavity}

\begin{figure*}[htb]
  \centering
  \includegraphics{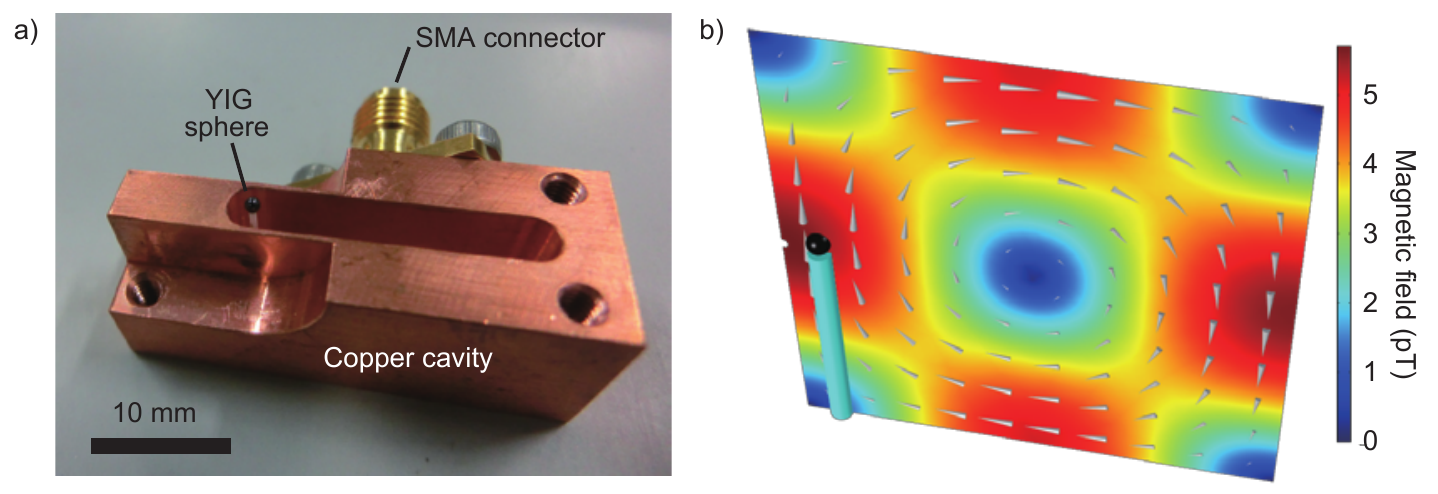}
  \caption{\label{fig:magcav_illust} Microwave cavity and YIG sphere. 
a) Photograph of a half of the microwave rectangular cavity made of oxygen-free copper. The cavity has dimensions of 22$\times$18$\times$3 mm and has the fundamental TE${}_{101}$-mode frequency of 10.565 GHz. An YIG sphere is mounted in the cavity with a support rod made of aluminum oxide. The sphere is glued to the rod oriented to the crystal axis $\langle 110\rangle$. b) Numerical simulation of the microwave magnetic-field distribution of the TE${}_{101}$ mode. The YIG sphere is located at the field maximum. }
\end{figure*}

In this section, we consider a sphere of ferromagnetic insulator embedded in a microwave cavity resonator. Related experiments have been reported recently by a few other groups~\cite{bib:Huebl13,bib:Zhang14,bib:Goryachev14}. We study the coupling between the Kittel mode and a single discretized microwave mode in a cavity in the quantum limit.

\subsection{Theory}

The coupling between linear-polarized microwave photons and the spin ensemble via the Zeeman effect is described by the Hamiltonian:
\begin{eqnarray}
  \hat{\mathcal{H}}_{\mathrm{int}} &=& g \mu_{\mathrm{B}} \sum_i  \hat{\mathbf{S}}_i \cdot  \mathbf{B}_0 (\mathbf{r}_i) \, (\hat{a}+\hat{a}^\dag) \nonumber \\
& = & g \mu_{\mathrm{B}} \sqrt{2s} 
\sum_{i} \sum_n
\mathbf{s}_n (\mathbf{r}_i) \, \frac{\hat{c}_n+\hat{c}_n^\dag}{2} \cdot  \mathbf{B}_0 (\mathbf{r}_i) \, (\hat{a}+\hat{a}^\dag), \nonumber \\ & & 
  \label{eq:magcav_hamil_int}
\end{eqnarray}
where $\mathbf{B}_0 (\mathbf{r}_i)$ is the linear-polarized microwave magnetic field in the cavity mode at the single photon level at the position $\mathbf{r}_i$ of each spin. In the second line, we replace the Heisenberg operator $\hat{\mathbf{S}}_i$ with the sum of magnon operators multiplied by their orthonormal mode functions $\mathbf{s}_n (\mathbf{r}_i)$.
Here, $n$ is an index of the modes.  

We then replace the sum over the spins with a volume integral and apply the rotating wave approximation, finally obtaining
\begin{equation}
  \hat{\mathcal{H}}_\mathrm{int} = 
      \frac{g \mu_{\mathrm{B}}}{2} \sqrt{2s}
      \sum_{n}
      \int_{V} d{\bf{r}} \, 
      \mathbf{s}_{n} (\mathbf{r}) \cdot \mathbf{B}_{0} (\mathbf{r}) 
      \left(   \hat{a}^\dag \hat{c}_{n}
       + \hat{a}\hat{c}_{n}^\dag 
      \right),
\end{equation}
where $V$ is the sample volume. 
For a cavity field spatially uniform within the sphere, we see from symmetry that the only mode with a finite coupling strength is the Kittel mode which has a spatially uniform function. 
In this case we obtain for $\mathbf{B}_0 \perp z$
\begin{eqnarray}
  \hat{\mathcal{H}}_\mathrm{int} 
    &=& \hbar g_{\mathrm{eff}} (\hat{a}^\dag \hat{c}+\hat{a}\hat{c}^\dag ), \\
    g_{\mathrm{eff}}& \equiv  &
    \frac{g \mu_{\mathrm{B}} B_0}{2\hbar} \sqrt{2sN}
    = \frac{\gamma B_0}{2} \sqrt{2sN}, 
     \label{eq:magcav_hamil_coupling}
\end{eqnarray}
where $\hat{c} = (1/\sqrt{2sN}) \sum_i {\hat{S}}_i^{+} $ is the annihilation operators of the Kittel mode, and $B_0 = |\mathbf{B}_0(\mathbf{r})|$.

\subsection{Yttrium iron garnet (YIG)}

In the following experiments, we use a single crystalline sphere of yttrium iron garnet ($\mathrm{Y_3Fe_5O_{12}}$; YIG) as a ferromagnetic sample. YIG is a celebrated ferromagnetic insulator~\cite{bib:Cherepanov93}, used for various microwave devices including filters and oscillators. The absence of conduction electrons leads to its small spin-wave relaxation rate, which also makes YIG very attractive in spintronics applications \cite{bib:Demokritov06,bib:Uchida10,bib:Kajiwara10}. Strictly speaking, YIG is a ferrimagnetic material, but all the spins in a unit cell practically precess in phase in the low energy limit, enabling us to treat it as ferromagnet. The net spin density $2sN/V$ in YIG is $2.1 \times 10^{22}\mu_\mathrm{B} \,\mathrm{cm}^{-3}$, orders of magnitude higher than the typical numbers, $10^{16}-10^{18}\mu_\mathrm{B}\, \mathrm{cm}^{-3}$, in the paramagnetic spin ensembles used in quantum memory experiments. Thus, we expect strong interaction of the spin excitations with an electromagnetic field.

\subsection{Experiment}

To accommodate an YIG sphere in the confined space, where only a single electromagnetic mode exists in a certain frequency range, we use a three-dimensional microwave cavity shown in Fig.~\ref{fig:magcav_illust}a. 
The picture shows a half cut of the cavity, and two pieces of them make a hollow cavity. 
The microwave magnetic-field distribution of the fundamental mode (rectangular TE${}_{101}$) is simulated by COMSOL Multiphysics\textsuperscript{\textregistered} (Fig.~\ref{fig:magcav_illust}b). 
A 0.5-mm-diameter YIG sphere is placed near the maximum of the magnetic field in order for obtaining the largest coupling strength and the uniformity of the field. 

We apply a static magnetic field of around 0.3~T by using a pair of neodymium permanent magnets and a $10^4$-turn superconducting coil.
They are connected in series using a magnetic yoke made of pure iron (Japanese Industrial Standard SUY-1). 
The static field is oriented to the $\langle 100\rangle$ crystal axis of the YIG sphere. 
The cavity has two SMA connectors for transmission spectroscopy. 
The center pins of the connectors are protruding into the cavity, such that their coupling strengths, $\kappa_{\mathrm{in}}/2\pi$ and $\kappa_{\mathrm{out}}/2\pi$, are about $0.5$~MHz. 
We use a weak probe microwave power of $-123$~dBm, which corresponds to the photon occupancy of 0.9 in the cavity mode. 
All the measurements are done in a dilution refrigerator: the ambient temperature at the sample stage is $10$~mK and the thermal photon/magnon occupancy at around 10~GHz is negligible. 

\begin{figure*}[htb]
  \centering
  \includegraphics{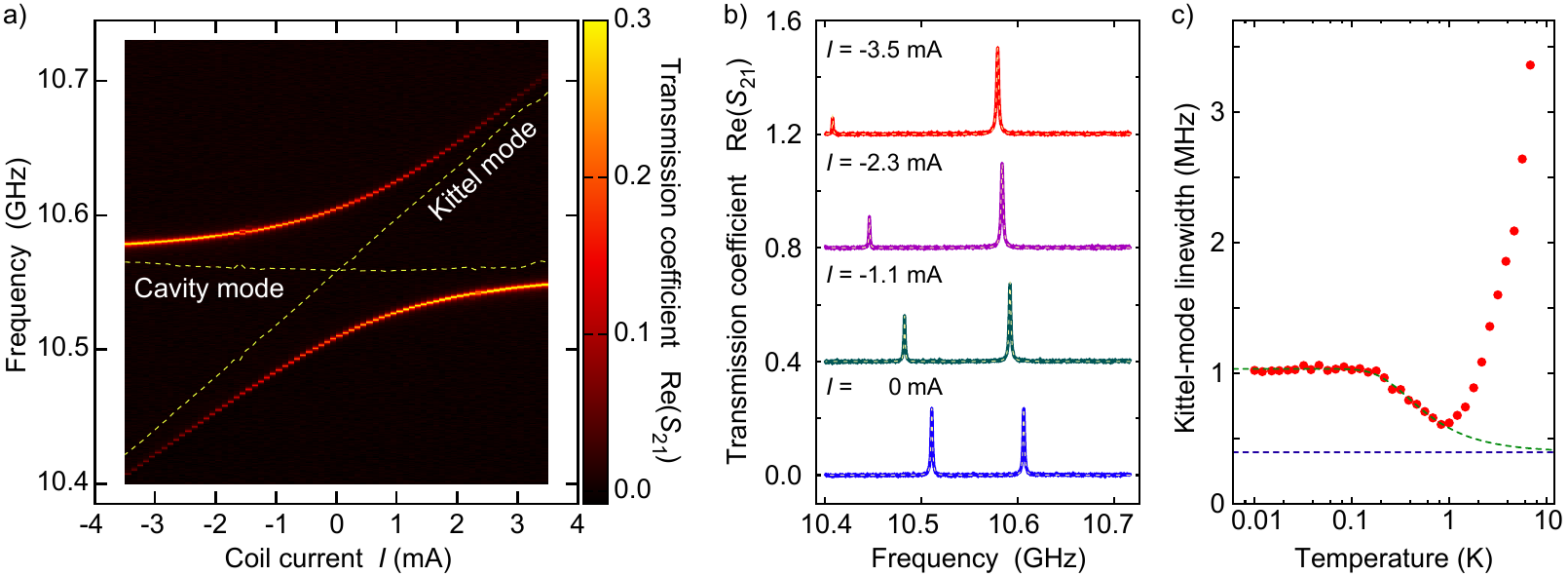}
  \caption{\label{fig:magcav_nms} Normal-mode splitting between the Kittel mode and the cavity mode TE${}_{101}$. a) Amplitude of the transmission Re($S_{21}$) through the cavity as a function of the probe microwave frequency and the static magnetic field represented in the current $I$ through the superconducting coil. The current $I$ is defined to be zero at the anticrossing. The horizontal and diagonal dashed lines show the TE${}_{101}$-mode and the Kittel-mode frequencies both obtained from the fitting. b) Cross sections at the static magnetic field corresponding to $I$ = $-3.5$, $-2.3$, $-1.1$, and $0$~mA. Solid curves are experimental data, and the dashed white lines are fitting curves based on the input-output theory. For clarity, the individual curves are offset vertically by 0.4 from bottom to top. c) Linewidth of the Kittel mode as a function of the temperature. The red dots show the linewidth obtained by fitting $S_{21}(\omega)$ measured at each temperature. The dashed line is the fitting curve to the temperature dependence below 1~K. The dashed blue line depicts the temperature-independent component of the fit. (Reprinted figure with permission from Y.~Tabuchi, \textit{et al.}, \textit{Phys. Rev. Lett.} \textbf{113}, 083603 (2014). Copyright (2014) by the American Physical Society.) }
\end{figure*}

We measure the transmission coefficient $S_{21}(\omega)$ of the cavity as a function of the probe frequency and the static magnetic field tuned by the bias current $I$ in the coil (Fig.~\ref{fig:magcav_nms}a). 
A large normal-mode splitting is observed, manifesting strong coupling between the Kittel mode and the cavity mode. 
Cross sections of the intensity plot are shown in Fig.~\ref{fig:magcav_nms}b. 
At the degeneracy point where the Kittel mode frequency coincides with the cavity frequency, we see two largely-separated peaks. 
The peaks indicate the hybridization of a Kittel-mode magnon and a cavity photon, i.e., formation of ``magnon-polaritons''. 
Their decay rates are apparently much smaller than the coupling energy. 

We quantify the coupling strength $g_m$, the cavity-mode decay rates $\kappa_\mathrm{in}+\kappa_\mathrm{out}+\kappa_\mathrm{int}$, and the Kittel-mode decay rate $\gamma_m$, based on the model Hamiltonian in Eq.~(\ref{eq:magcav_hamil_coupling}).
Here, $\kappa_\mathrm{in}$, $\kappa_\mathrm{out}$, and  $\kappa_\mathrm{int}$ are the cavity decay rates through the input and output ports and the internal loss channel, respectively.
Using the input-output theory, we derive the transmission coefficient $S_{21}(\omega)$ as:
\begin{equation}
  S_{21}(\omega) = \frac{\sqrt{\kappa_{\mathrm{in}}\kappa_{\mathrm{out}} }} 
  {i(\omega-\omega_{\textrm{\scriptsize c}})
  -\frac{\kappa_{\mathrm{in}}+\kappa_{\mathrm{out}}+\kappa_{\mathrm{int}}}{2} 
 +\frac{|g_{\mathrm{m}}|^2}{i(\omega-\omega_{\mathrm{m}})-\gamma_{\mathrm{m}}}}.
  \label{eq:magcav_spara}
\end{equation}
The fitting curves, shown as the white dashed lines in Fig.~\ref{fig:magcav_nms}b,  well reproduce the experimental data. 
From the fitting, we obtain  $g_m/2\pi = 47$~MHz, $(\kappa_{\mathrm{in}}+\kappa_{\mathrm{out}}+\kappa_{\mathrm{int}})/2\pi = 2.7$~MHz, and $\gamma_{\mathrm{m}}/2\pi = 1.1$~MHz. 
We find our magnon-cavity hybrid system deep in the strong coupling regime, $g_{\mathrm{m}} \gg \kappa, \gamma$, even in the quantum limit where the average photon/magnon number is less than one. 

\subsection{Coupling strength}

The obtained coupling strength $g_{\mathrm{m}}/2\pi$ of $47$~MHz results from the $\sqrt{N}$-enhancement according to Eq.~(\ref{eq:magcav_hamil_coupling}). 
Given that the $0.5$-mm-diameter sphere contains $1.4 \times 10^{18}$ net spins, the single spin coupling strength is estimated to be $40$~mHz. 

In designing coupling strengths for various applications, it is worth estimating the strength with numerical simulation. Figure~\ref{fig:magcav_illust}b shows the magnetic field distribution $B_0$ at the single photon level of the TE${}_{101}$ mode. The coupling strength can be easily calculated by the relation $g_0 = \gamma B_0/2$. The simulated value of $B_0 = 5.5$~pT/photon at the sample gives $g_0/2\pi = 38.5$~mHz, which agrees well with the experiment. 

\subsection{Magnon linewidth}

Little has been known about the linewidth of the Kittel mode in the temperature range attainable in a dilution refrigerator. We measure the temperature dependence of the resonance linewidth below 1~K (Fig.~\ref{fig:magcav_nms}c) and  observe a peculiar behavior below 1~K; the linewidth is broadened as temperature decreases. 

The fitting curve based on the TLS model, as indicated with the green dashed line in Fig.~\ref{fig:magcav_nms}c, well agrees with the experimental data below 1~K. Note that the Kittel-mode frequency $\omega_{\mathrm{m}}$ is used as a fixed parameter in the temperature-dependent term proportional to $\tanh(\hbar \omega_{\mathrm{m}} / 2 k_{\mathrm{B}} T)$. We also assume the presence of a temperature-independent contribution in the fitting. The parameters obtained imply that among the linewidth at the lowest temperature a fraction of $0.63$~MHz is attributed to TLSs, and the other $0.39$~MHz to surface scattering. The microscopic origin of the TLSs remains to be understood. The linewidth broadening above 1 K is ascribed to the slow-relaxation mechanism caused by the magnetic impurities~\cite{bib:Teale62}. 

An additional signature of the effect of the TLSs is found in the power dependence of the linewidth.
Strong microwave drive saturates the TLSs coupled to the Kittel mode, resulting in the narrowing of the linewidth.
Similar phenomena have been observed in superconducting microwave resonators interacting with TLSs~\cite{bib:OConnell08,bib:Jiansong08}.
We confirmed experimentally that the Kittel-mode linewidth indeed became narrower at higher power of the drive.
The power level causing the narrowing should be related with the dipole strengths and the relaxation rates of the TLSs. 
Comprehensive and systematic analyses of the linewidth with respect to the diameter and the quality of the spheres, the crystal orientations to the static field, and the Kittel mode frequency are also awaited. 

\section{Coupling with a superconducting qubit}\label{qubit}

\begin{figure*}[htb]
  \centering
  \includegraphics{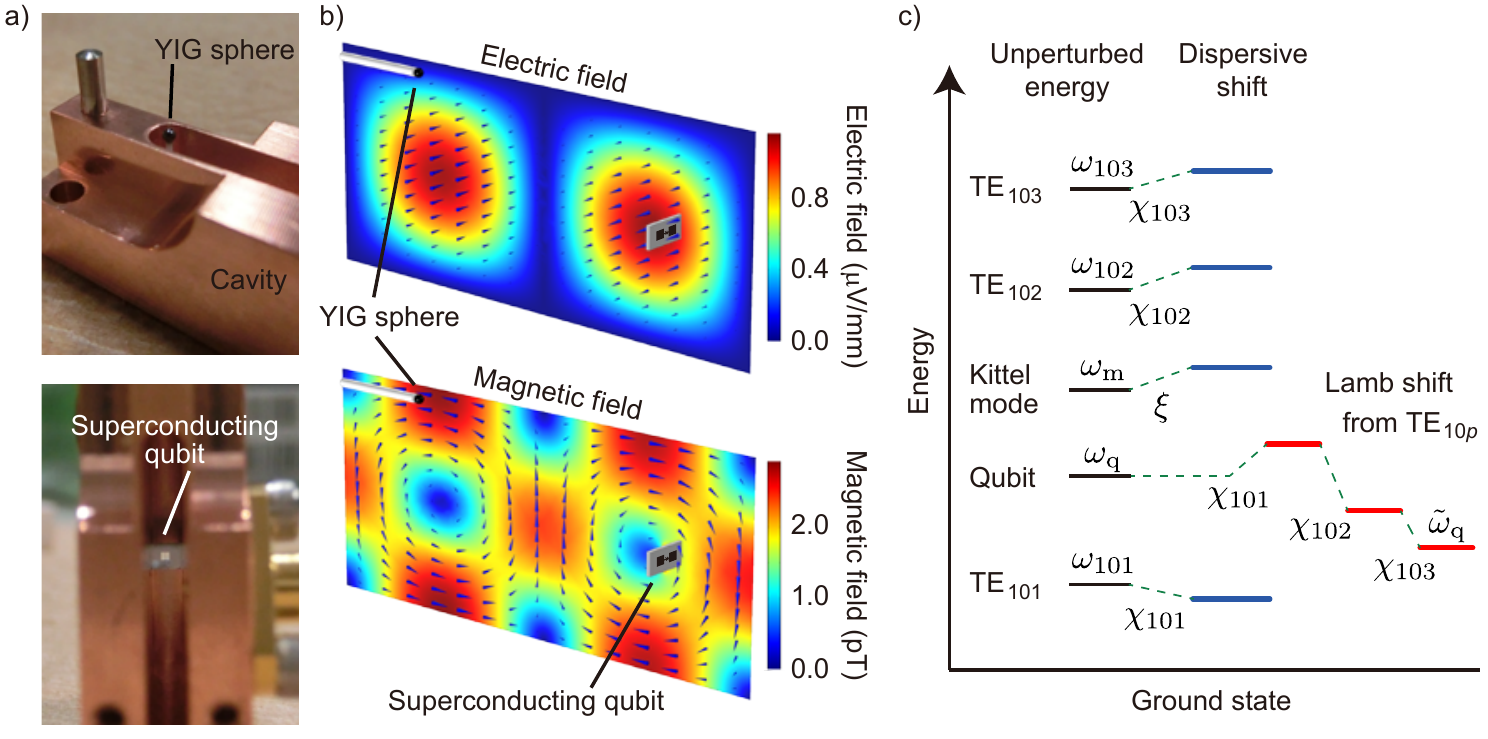}
  \caption{\label{fig:qmag_outline} Qubit-cavity-YIG hybrid system. a) Photographs of an YIG sphere and a superconducting qubit mounted in an extended cavity resonator. The YIG sphere is mounted in one side of the cavity. On the other side, the superconducting transmon qubit fabricated on a silicon chip is seen in the deep cavity trench. The cavity with dimensions of $25\times 3\times 53$~mm supports the lowest-frequency (TE${}_{101}$, TE${}_{102}$, and TE${}_{103}$) modes with the frequencies of $6.987$~GHz, $8.488$~GHz, and $10.461$~GHz, respectively. b) Numerical simulation of the TE${}_{102}$ mode. Intensities of the microwave electric and magnetic fields are shown, respectively; the blue arrows indicate the directions of the fields. The qubit chip is placed near the electric field antinode. The qubit with a millimeter-sized dipole antenna strongly couples with the electric field. The YIG crystal is placed where the magnetic field is large. The color scales indicate field intensities at the single photon level. c) Schematic energy diagram of the qubit-cavity-YIG hybrid system. The left shows unperturbed (bare) frequencies of the qubit, the cavity and the Kittel modes. The cavity and the Kittel modes are subject to frequency shifts due to the coupling with the qubit, as indicated in the middle of the diagram. The cavity modes induce the Lamb shift of the qubit, as depicted on the right-hand side. }
\end{figure*}

In the previous section we demonstrated coherent coupling of magnons to the cavity mode and investigated the linewidth in the quantum limit. However, that was a linear coupling between two harmonic oscillator modes. Although we observed the normal-mode splitting at the quantum limit near the ground state, there was no significant difference from the one we see in the classical limit, e.g., at room temperature and with much stronger probe microwave. The correspondence principle makes it difficult to distinguish quantum and classical behaviors in such a linear system. Nonlinearity, or aharmonicity, is required for a clear demonstration of the quantum behavior. Thus, in this section, we move one step further by introducing a well-controllable two-level system, i.e., a superconducting qubit. 

\subsection{Theory}

In order to implement a coupling between a superconducting qubit and a magnon in YIG, we exploit the cavity quantum electrodynamics architecture.
In the scheme illustrated below, two heterogeneous systems --- the qubit and the Kittel mode --- are linked through electromagnetic fields in a microwave cavity resonator. 

We use an elongated cavity shown in Fig.~\ref{fig:qmag_outline}a, which simultaneously accommodates a qubit and an YIG sphere. 
The 5-cm-long cavity has eigenmodes of TE${}_{10p}$ $(p=1,2,\cdots)$ as the lowest-frequency modes. 
The resonant frequencies $\omega_{10p}$ are determined by the width $W$ and length $L$ of the cavity, and denoted as:
\begin{equation}
  \omega_{10p} = \frac{\pi}{2} c_0 
  \sqrt{
    \left(\frac{1}{W}\right)^2
    +\left(\frac{p}{L}\right)^2},
\end{equation}
where $c_0$ is the speed of light. 
In Fig.~\ref{fig:qmag_outline}b the simulated electric and magnetic field distributions of the TE${}_{102}$ mode are shown. 

We use a transmon qubit which has two aluminum pads and a Josephson junction bridging them. 
The qubit chip is embedded in the cavity as seen in the bottom picture of Fig.~\ref{fig:qmag_outline}a. 
The submillimeter-sized dipole antenna consisting of the two pads electrically couples to the microwave field in the cavity. 
The qubit is placed where the electric field is close to the maximum. 
On the other hand, the Kittel mode of the YIG sphere magnetically couples to the same cavity mode. 
The corresponding system Hamiltonian is written as:
\begin{eqnarray}
  \hat{\cal H}_{\mathrm{sys}}/\hbar &=& \sum_{p} \omega_{10p}\, \hat{a}_{p}^\dagger \hat{a}_{p}
  + \left[(\omega_{\mathrm{q}}-\alpha/2)\,\hat{b}^\dagger \hat{b} 
  + (\alpha/2)\,(\hat{b}^\dagger \hat{b})^2\right] \nonumber \\
  &+& \omega_{\mathrm{m}}\,\hat{c}^\dagger \hat{c}, 
\end{eqnarray}
where $\omega_{10p}$ is the resonant frequency of TE${}_{10p}$, $\omega_{\mathrm{q}}$ and $\omega_{\mathrm{m}}$ are the qubit and the Kittel mode frequencies, and $\hat{a}_{p}$, $\hat{b} \equiv \sum_{l \geq 0} \sqrt{l+1}\,|l\rangle_{\mathrm{q}} {}_{\mathrm{q}}\langle l+1|$ and $\hat{c}$ are annihilation operators for the cavity mode, the qubit, and the Kittel mode, respectively. 
The transmon qubit is represented as an anharmonic oscillator with an anharmonicity of $\alpha < 0$. 
The lowering operator for the qubit subspace is defined as $\sigma^{-} = [\sigma^{+}]^\dagger = \hat{\cal P}\,\hat{b}$, where $\hat{\cal P} = |0\rangle \langle0|+|1\rangle \langle1|$ is a projection operator. 
The annihilation operator of the Kittel mode $\hat{c} = (1/\sqrt{2sN})\,\sum_i \hat{S}_i^{+}$ stems from the Holstein-Primakoff transformation. 
The interaction Hamiltonian is written as:
\begin{widetext}
\begin{eqnarray}
  \hat{\cal H}_{\mathrm{int}}/\hbar 
    &=& \left(\sum_p g_{\mathrm{q},10p}\,\hat{b}^\dagger \hat{a}_{\mathrm{p}} 
       + h.c. \right) 
   +  \left( \sum_p g_{\mathrm{m},10p}\,\hat{c}^{\dagger} \hat{a}_{\mathrm{p}}
        + h.c. \right), \label{eq:qmag_int}
\end{eqnarray}
\end{widetext}
where $g_{\mathrm{q},10p}$ and $g_{\mathrm{m},10p}$ are the coupling strengths of the cavity TE${}_{10p}$ mode to the qubit and the Kittel mode, respectively. 
Note that the direct interaction between the qubit and the Kittel mode is negligible.
In the large detuning regime where $g_{\mathrm{q},10p}, g_{\mathrm{m},10p} < |\omega_{\mathrm{q}}-\omega_{\mathrm{m}}| \ll |\omega_{10p} - \omega_{\mathrm{q}}|, |\omega_{10p} - \omega_{\mathrm{m}}|$, the interaction Hamiltonian can be rewritten in the rotating frame of the system Hamiltonian under a perturbative treatment up to the first order~\cite{bib:Scholz10}, and expressed as:
\begin{widetext}
\begin{eqnarray}
  {\tilde{\cal H}}_{\mathrm{int}}/\hbar &=& 
  \sum_{p} 
    \frac{g_{\mathrm{m},10p}^2}{\omega_{10p}-\omega_{\mathrm{m}}} 
    \hat{a}_{p}^\dagger \hat{a}_{p} 
 -\sum_{p}
    \sum_{l>0}
    \lambda_{10p}^{(l)}\,|l\rangle_{\mathrm{q}} {}_{\mathrm{q}}\langle l|
 -\sum_{p}
    \frac{g_{\mathrm{m},10p}^2}{\omega_{10p}-\omega_{\mathrm{m}}}\,
    \hat{c}^\dagger \hat{c} \nonumber \\
  &+& 
  \sum_{p} 
    \left(
      \chi_{10p}|0\rangle_{\mathrm{q}} {}_{\mathrm{q}}\langle 0|
      + \sum_{l>0}\chi_{10p}^{(l)}
             |l\rangle_{\mathrm{q}} {}_{\mathrm{q}}\langle l|\right)
      \hat{a}_{10p}^\dagger \hat{a}_{10p} 
 -\sum_{p} \frac{1}{N} 
    \frac{2g_{\mathrm{m},10p}^2}{\omega_{10p}-\omega_{\mathrm{m}}}\,
    \hat{a}_{10p}^\dagger\hat{a}_{10p}\,\hat{c}^\dagger\hat{c}, 
  \label{eq:qmag_dispersive}
\end{eqnarray}
\end{widetext}
where $\chi_{10p} = |g_{\mathrm{q},10p}|^2/(\omega_{10p} - \omega_{\mathrm{q}})$, 
$\lambda_{10p}^{(l)} = l\,|g_{\mathrm{q},10p}|^2/[\omega_{10p}-\omega_{\mathrm{q}}-(l-1)\alpha]$, $\chi_{10p}^{(l)} = |g_{\mathrm{q},10p}|^2[(l+1)/(\omega_{10p}-\omega_{\mathrm{q}}-l\,\alpha) - l/(\omega_{10p}-\omega_{\mathrm{q}}-(l-1)\,\alpha)]$. 
The first and the third terms are dispersive shifts due to the coupling between the cavity modes and the Kittel mode. 
The second term indicates the Lamb shift of the qubit frequency arising from the coupling to the cavity mode. 
The Lamb shift for the first excited states $\lambda_{10p}^{(1)}$ coincides with $\chi_{10p}$. The fourth term shows the qubit-state-dependent cavity frequency shift or the photon-number-dependent qubit frequency shift. It is worth noting that the last term indicates the static interaction between the cavity modes and the Kittel mode, which originates from the fact that the spin ensemble is not a perfect bosonic system. 
Because of the factor $1/N$, however, for large $N$ such effect is not observable with usual experimental parameters. 
The major energy shifts are summarized in Fig.~\ref{fig:qmag_outline}c. 
In the regime where $g_{\mathrm{q},10p}g_{\mathrm{m},10p}/\sqrt{|\omega_{10p}-\omega_{\mathrm{q}}|\,|\omega_{10p}-\omega_{\mathrm{m}}|} < |\omega_{\mathrm{q}}-\omega_{\mathrm{m}}| \ll |\omega_{10p}-\omega_{\mathrm{q}}|,|\omega_{10p}-\omega_{\mathrm{m}}|$, there is a qubit-state-dependent frequency shift $\xi$ of the Kittel mode~\cite{bib:Tabuchi15}. 
Although it appears only in the third-order perturbative treatment, the shift is still observable when the Kittel-mode and the qubit frequencies are close enough to meet the frequency condition. 
Such coupling is usable for counting the magnon number in the Kittel mode via a Ramsey interferometry using the qubit, for example. 

The coupling between the qubit and the Kittel mode is mediated by the cavity modes TE${}_{10p}$ when the qubit and the Kittel-mode frequencies are degenerate with each other and detuned from the cavity so that $|\omega_{\mathrm{q}} - \omega_{\mathrm{m}} | \ll g_{\mathrm{q},10p}, g_{\mathrm{m},10p} \ll |\omega_{10p}-\omega_{\mathrm{q}}| \simeq |\omega_{10p}-\omega_{\mathrm{m}}|$~\cite{bib:Imamoglu09}. 
We rewrite the interaction Hamiltonian in Eq.~(\ref{eq:qmag_int}) in the corresponding rotating frame by adiabatically eliminating the cavity modes:
\begin{eqnarray}
  {\tilde{\cal H}}_{\mathrm{int}}/\hbar &=& 
   g_{\mathrm{q-m}}\,\hat{\sigma}^{-}\hat{c}^\dagger + h.c.,
   \label{eq:qmag_coupling}
\end{eqnarray}
where $g_{\mathrm{q-m}} = \sum_{p} g_{\mathrm{m},10p}\,g_{\mathrm{q},10p} / (\omega_{10p}-\omega_{\mathrm{q}})$. It is interpreted as that the qubit and the Kittel mode exchange their energy quanta, i.e., the qubit excitation and a magnon, through the virtual excitation of the cavity mode 
TE${}_{10p}$. 

\subsection{Experiment}
 
\begin{figure*}[htb]
  \centering
  \includegraphics{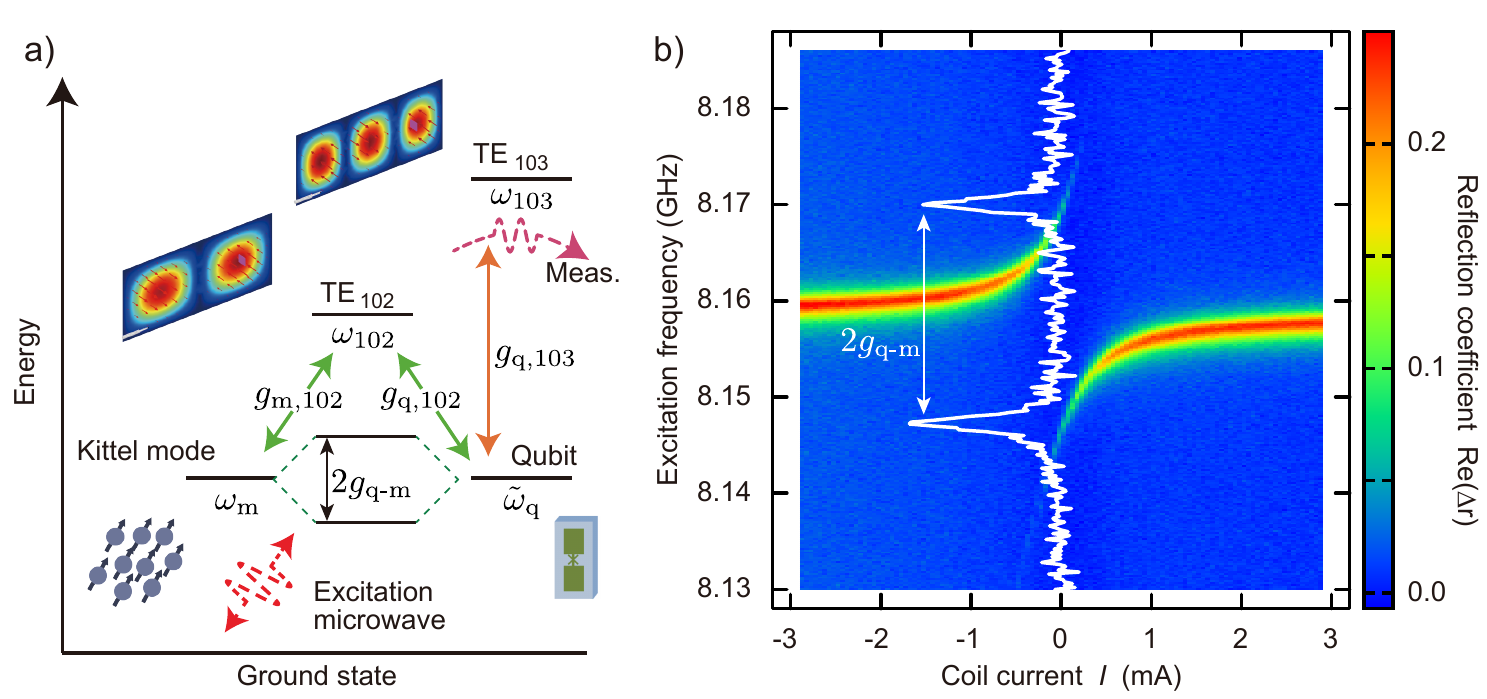}
  \caption{\label{fig:qmag_illust} Magnon-qubit hybridization. 
a) Energy level diagram illustrating the coupling scheme. The superconducting transmon qubit is detuned from the TE${}_{102}$ cavity mode by $-178$~MHz, and the Kittel mode is almost degenerate with the qubit. 
The couplings of the TE${}_{102}$ mode with the qubit and the Kittel mode induce their interaction via the virtual-photon excitation; 
the qubit-magnon degeneracy is lifted by $2 g_{\mathrm{q-m}}$. 
The TE${}_{103}$ cavity mode is dispersively coupled to the qubit so that its resonant frequency depends on the qubit state. 
b) Magnon-vacuum-induced Rabi splitting. 
The change in the real part of the reflection coefficient at the TE${}_{103}$ frequency, $\mathrm{Re}(\Delta r)$, is plotted as a function of the qubit excitation frequency and the static magnetic field which is locally applied to the YIG sphere. 
The static field is represented in the coil current $I$, which is defined to be zero at the anticrossing. 
The white solid curve shows the cross section at $I$ = $0$~mA. The powers used for probing the TE${}_{103}$ mode and for exciting the qubit and the Kittel mode are $-135$~dBm and $-130$~dBm, respectively. Note that the measurement here was done with the same sample set as in Ref.~\cite{bib:Tabuchi15}. By readjusting the alignment of the YIG sphere, we were able to remove the effect of other magnetostatic modes previously observed.}
\end{figure*}

Here we focus on the roles played by the cavity modes TE${}_{102}$ and TE${}_{103}$ (Fig.~\ref{fig:qmag_illust}a). 
The cavity mode TE${}_{102}$ is utilized for mediating the virtual-photon interaction between the qubit and the Kittel mode, while the cavity mode TE${}_{103}$ is used for the qubit measurement. 
Although the Kittel mode also couples to the TE${}_{103}$ mode, we can perform a selective qubit measurement via the mode because the magnon-number-dependent shift of the last term in Eq.~(\ref{eq:qmag_dispersive}) is sufficiently small and safely ignored. 
We introduce probe and excitation microwaves into the cavity through an SMA port. 
The external coupling strengths and internal losses of each cavity mode are 
$\kappa_{\mathrm{103}}/2\pi$ = $2.75$~MHz, %
$\kappa_{\mathrm{103,int}}/2\pi$ = $1.26$~MHz, %
$\kappa_{\mathrm{102}}/2\pi$ = $0.55$~MHz, and %
$\kappa_{\mathrm{103,int}}/2\pi$ = $1.73$~MHz, respectively. 

A 0.5-mm-diameter YIG sphere is placed where the microwave magnetic field of the cavity mode TE${}_{102}$ is substantial, as indicated in Fig.~\ref{fig:qmag_outline}b. 
The field is parallel to the $\langle 110 \rangle$ crystal axis of the YIG sphere. 
We also apply a static magnetic field locally to the sphere in parallel with its $\langle 100 \rangle$ crystal axis. 
A doubly-layered magnetic shield protects a superconducting qubit from the stray magnetic field which is approximately a few tens of gauss at the qubit position. 
The coupling strength $g_{\mathrm{m,102}}$ of the Kittel mode to the cavity mode TE${}_{102}$ and the decay rate $\gamma_{\mathrm{m}}$ of the Kittel mode are obtained from the anticrossing spectrum in the reflection measurement: $g_{\mathrm{m,102}}/2\pi$ = $21$~MHz and $\gamma_{\mathrm{m}}/2\pi$=$1.8$~MHz. 
The transmon qubit has the dressed resonant frequency $\tilde{\omega}_{\mathrm{q}}/2\pi = (\omega_{\mathrm{q}} - \sum_{\mathrm{q}} \chi_{10p})/2\pi$ of $8.158$~GHz and the anharmonicity $\alpha/2\pi$ of $-158$~MHz. 
A separate time-domain measurement shows that the energy relaxation time $T_{1} = 273$~ns of the qubit is almost at the Purcell limit where $1/T_1 = \sum_{p}\chi_{10p} (\kappa_{10p}+\kappa_{10p,\mathrm{int}})/(\omega_{10p}-\omega_{\mathrm{q}})$. 
We deduce the Lamb shift $\lambda_{10p}^{(1)} = \chi_{10p}$ from the frequency difference between the bare- and the dressed-cavity frequencies, which is accurate up to the first-order perturbative treatment; they are evaluated to be $\chi_{\mathrm{102}}/2\pi = 75$~MHz and $\chi_{\mathrm{103}}/2\pi = 9$~MHz. 
The coupling strengths of the qubit to the cavity modes are $g_{\mathrm{q,102}}/2\pi$ = $117$~MHz and  $g_{\mathrm{q,103}}/2\pi$ = $141$~MHz. The qubit-state-dependent cavity shift $(\chi_{103}^{(1)}-\chi_{103})/2\pi$ of $1.2$~MHz is used for the qubit spectroscopy. 
The qubit linewidth $\gamma_{\mathrm{q}}/2\pi$ is 2.0~MHz. 

Figure~\ref{fig:qmag_illust}a illustrates the measurement scheme of the hybridized qubit-magnon modes. 
At the degeneracy point where the Kittel-mode frequency coincides with the qubit frequency, the virtual excitation of the cavity mode TE${}_{102}$ mediates interaction between them as expressed in Eq.~(\ref{eq:qmag_coupling}), so that the qubit frequency splits into two branches. 
When we apply a resonant microwave to the branches, they are excited and detected as a phase shift in the reflection coefficient of the cavity probe microwave at the TE${}_{103}$ frequency, given that each branch has a fraction of the qubit wave function. 
In this scheme which keeps the cavity mode TE${}_{102}$ empty, unwanted dephasing caused by photon number fluctuations in the ``coupler'' mode can be avoided. 

\subsection{Result and discussion}

Figure~\ref{fig:qmag_illust}b shows the reflection coefficient at the TE${}_{103}$ frequency as a function of the excitation frequency and the coil current. 
The pronounced splitting of the qubit frequency at the degeneracy point manifests magnon-vacuum-induced Rabi splitting, which indicates coherent coupling between the qubit and the Kittel mode. 
The obtained coupling strength $g_{\mathrm{q-m}}/2\pi$ is 11.4~MHz, which is much larger than the qubit decay rate $\gamma_{\mathrm{q}}$ and the Kittel-mode decay rate $\gamma_{\mathrm{m}}$. 
Thus the hybridized qubit-magnon system also stands deep in the strong coupling regime where $g_{\mathrm{q-m}} \gg \gamma_{\mathrm{q}}, \gamma_{\mathrm{m}}$.

By using Eq.~(\ref{eq:qmag_coupling}), we estimate the coupling strength for a detuning $\omega_{102}-\omega_{\mathrm{q}} = \omega_{102} - \tilde{\omega}_{\mathrm{q}} - \sum_{p=1,2,3} \lambda^{(1)}_{10p}$ = $183$~MHz and obtain $13.4$~MHz. 
The discrepancy of 17\% between the calculated and the measured strengths is ascribed to the first-order approximation of the Hamiltonian; the value  $|g_{\mathrm{q,102}}/(\omega_{102}-\omega_{\mathrm{q}})|$ $\sim$ $0.6$ should be far less than unity for a convergence of the series. 
Another possible factor is the estimation error in the bare qubit frequency $\omega_{\mathrm{q}}$.
Because all the cavity modes contribute to the Lamb shift of the qubit, it is not straightforward to accurately estimate the shift in a multi-mode cavity, especially when the modes are crowded. 

\section{Conclusions and outlook}

\label{conclusion}

We demonstrated coherent coupling between a single magnon excitation in a ferromagnet and a microwave photon in a cavity as well as a superconducting qubit. 
It was proven that the uniform spin precession mode, i.e., the Kittel mode, of a millimeter-scale ferromagnetic sphere behaves quantum mechanically:
the qubit showed a Rabi splitting induced by the vacuum fluctuations of the Kittel mode.

The technique developed here exploits the advanced circuit-QED and microwave quantum optics technologies based on superconducting qubits.
It enables generation and characterization of non-classical states of magnons and thus opens a new field of quantum magnonics.
It is also considered to be the ultimate limit of spintronics. 
We expect that further studies will reveal the dissipation mechanisms in the single magnon regime.

The coherent transfer of the qubit state to the magnon mode suggests a possible link to quantum information networks in optics. In contrast to superconducting circuits, spins in insulating crystals can interact coherently with light, as demonstrated in quantum memory experiments in the optical domain~\cite{bib:Jevon05,bib:Hammerer10}. In those experiments, paramagnetic spin ensembles have been used. It is of great interest to look into coherent coupling between magnons in ferromagnet and light.




\section*{Acknowledgements}
The authors acknowledge P.-M. Billangeon for fabricating the transmon qubit. 
This work was partly supported by the Project for Developing Innovation System of MEXT, JSPS KAKENHI (Grant Number 26600071, 26220601), the Murata Science Foundation, Research Foundation for Opto-Science and Technology, and NICT.

\bibliography{tabuchi}

\begin{thebibliography}{42}%
\makeatletter
\providecommand \@ifxundefined [1]{%
 \@ifx{#1\undefined}
}%
\providecommand \@ifnum [1]{%
 \ifnum #1\expandafter \@firstoftwo
 \else \expandafter \@secondoftwo
 \fi
}%
\providecommand \@ifx [1]{%
 \ifx #1\expandafter \@firstoftwo
 \else \expandafter \@secondoftwo
 \fi
}%
\providecommand \natexlab [1]{#1}%
\providecommand \enquote  [1]{``#1''}%
\providecommand \bibnamefont  [1]{#1}%
\providecommand \bibfnamefont [1]{#1}%
\providecommand \citenamefont [1]{#1}%
\providecommand \href@noop [0]{\@secondoftwo}%
\providecommand \href [0]{\begingroup \@sanitize@url \@href}%
\providecommand \@href[1]{\@@startlink{#1}\@@href}%
\providecommand \@@href[1]{\endgroup#1\@@endlink}%
\providecommand \@sanitize@url [0]{\catcode `\\12\catcode `\$12\catcode
  `\&12\catcode `\#12\catcode `\^12\catcode `\_12\catcode `\%12\relax}%
\providecommand \@@startlink[1]{}%
\providecommand \@@endlink[0]{}%
\providecommand \url  [0]{\begingroup\@sanitize@url \@url }%
\providecommand \@url [1]{\endgroup\@href {#1}{\urlprefix }}%
\providecommand \urlprefix  [0]{URL }%
\providecommand \Eprint [0]{\href }%
\providecommand \doibase [0]{http://dx.doi.org/}%
\providecommand \selectlanguage [0]{\@gobble}%
\providecommand \bibinfo  [0]{\@secondoftwo}%
\providecommand \bibfield  [0]{\@secondoftwo}%
\providecommand \translation [1]{[#1]}%
\providecommand \BibitemOpen [0]{}%
\providecommand \bibitemStop [0]{}%
\providecommand \bibitemNoStop [0]{.\EOS\space}%
\providecommand \EOS [0]{\spacefactor3000\relax}%
\providecommand \BibitemShut  [1]{\csname bibitem#1\endcsname}%
\let\auto@bib@innerbib\@empty
\bibitem [{\citenamefont {Schoelkopf}\ and\ \citenamefont
  {Girvin}(2008)}]{bib:Schoelkopf08}%
  \BibitemOpen
  \bibfield  {author} {\bibinfo {author} {\bibfnamefont {R.~J.}\ \bibnamefont
  {Schoelkopf}}\ and\ \bibinfo {author} {\bibfnamefont {S.~M.}\ \bibnamefont
  {Girvin}},\ }\href@noop {} {\bibfield  {journal} {\bibinfo  {journal}
  {Nature}\ }\textbf {\bibinfo {volume} {451}},\ \bibinfo {pages} {664}
  (\bibinfo {year} {2008})}\BibitemShut {NoStop}%
\bibitem [{\citenamefont {You}\ and\ \citenamefont {Nori}(2011)}]{bib:You11}%
  \BibitemOpen
  \bibfield  {author} {\bibinfo {author} {\bibfnamefont {J.~Q.}\ \bibnamefont
  {You}}\ and\ \bibinfo {author} {\bibfnamefont {F.}~\bibnamefont {Nori}},\
  }\href {\doibase 10.1038/nature10122} {\bibfield  {journal} {\bibinfo
  {journal} {Nature}\ }\textbf {\bibinfo {volume} {474}},\ \bibinfo {pages}
  {589} (\bibinfo {year} {2011})}\BibitemShut {NoStop}%
\bibitem [{\citenamefont {Devoret}\ and\ \citenamefont
  {Schoelkopf}(2013)}]{bib:Devoret13}%
  \BibitemOpen
  \bibfield  {author} {\bibinfo {author} {\bibfnamefont {M.~H.}\ \bibnamefont
  {Devoret}}\ and\ \bibinfo {author} {\bibfnamefont {R.~J.}\ \bibnamefont
  {Schoelkopf}},\ }\href {\doibase 10.1126/science.1231930} {\bibfield
  {journal} {\bibinfo  {journal} {Science}\ }\textbf {\bibinfo {volume}
  {339}},\ \bibinfo {pages} {1169} (\bibinfo {year} {2013})}\BibitemShut
  {NoStop}%
\bibitem [{\citenamefont {Riste}\ \emph {et~al.}(2013)\citenamefont {Riste},
  \citenamefont {Dukalski}, \citenamefont {Watson}, \citenamefont {de~Lange},
  \citenamefont {Tiggelman}, \citenamefont {Blanter}, \citenamefont {Lehnert},
  \citenamefont {Schouten},\ and\ \citenamefont {DiCarlo}}]{bib:Riste13}%
  \BibitemOpen
  \bibfield  {author} {\bibinfo {author} {\bibfnamefont {D.}~\bibnamefont
  {Riste}}, \bibinfo {author} {\bibfnamefont {M.}~\bibnamefont {Dukalski}},
  \bibinfo {author} {\bibfnamefont {C.~A.}\ \bibnamefont {Watson}}, \bibinfo
  {author} {\bibfnamefont {G.}~\bibnamefont {de~Lange}}, \bibinfo {author}
  {\bibfnamefont {M.~J.}\ \bibnamefont {Tiggelman}}, \bibinfo {author}
  {\bibfnamefont {Y.~M.}\ \bibnamefont {Blanter}}, \bibinfo {author}
  {\bibfnamefont {K.~W.}\ \bibnamefont {Lehnert}}, \bibinfo {author}
  {\bibfnamefont {R.~N.}\ \bibnamefont {Schouten}}, \ and\ \bibinfo {author}
  {\bibfnamefont {L.}~\bibnamefont {DiCarlo}},\ }\href {\doibase 10.1038/nature12513} {\bibfield  {journal} {\bibinfo  {journal} {Nature}\
  }\textbf {\bibinfo {volume} {502}},\ \bibinfo {pages} {350} (\bibinfo {year}
  {2013})}\BibitemShut {NoStop}%
\bibitem [{\citenamefont {Weber}\ \emph {et~al.}(2014)\citenamefont {Weber},
  \citenamefont {Chantasri}, \citenamefont {Dressel}, \citenamefont {Jordan},
  \citenamefont {Murch},\ and\ \citenamefont {Siddiqi}}]{bib:Weber14}%
  \BibitemOpen
  \bibfield  {author} {\bibinfo {author} {\bibfnamefont {S.~J.}\ \bibnamefont
  {Weber}}, \bibinfo {author} {\bibfnamefont {A.}~\bibnamefont {Chantasri}},
  \bibinfo {author} {\bibfnamefont {J.}~\bibnamefont {Dressel}}, \bibinfo
  {author} {\bibfnamefont {A.~N.}\ \bibnamefont {Jordan}}, \bibinfo {author}
  {\bibfnamefont {K.~W.}\ \bibnamefont {Murch}}, \ and\ \bibinfo {author}
  {\bibfnamefont {I.}~\bibnamefont {Siddiqi}},\ }\href {\doibase 10.1038/nature13559} {\bibfield  {journal} {\bibinfo  {journal} {Nature}\
  }\textbf {\bibinfo {volume} {511}},\ \bibinfo {pages} {570} (\bibinfo {year}
  {2014})}\BibitemShut {NoStop}%
\bibitem [{\citenamefont {Kelly}\ \emph {et~al.}(2015)\citenamefont {Kelly},
  \citenamefont {Barends}, \citenamefont {Fowler}, \citenamefont {Megrant},
  \citenamefont {Jeffrey}, \citenamefont {White}, \citenamefont {Sank},
  \citenamefont {Mutus}, \citenamefont {Campbell}, \citenamefont {Chen},
  \citenamefont {Chen}, \citenamefont {Chiaro}, \citenamefont {Dunsworth},
  \citenamefont {Hoi}, \citenamefont {Neill}, \citenamefont {O'Malley},
  \citenamefont {Quintana}, \citenamefont {Roushan}, \citenamefont
  {Vainsencher}, \citenamefont {Wenner}, \citenamefont {Cleland},\ and\
  \citenamefont {Martinis}}]{bib:Kelly15}%
  \BibitemOpen
  \bibfield  {author} {\bibinfo {author} {\bibfnamefont {J.}~\bibnamefont
  {Kelly}}, \bibinfo {author} {\bibfnamefont {R.}~\bibnamefont {Barends}},
  \bibinfo {author} {\bibfnamefont {A.~G.}\ \bibnamefont {Fowler}}, \bibinfo
  {author} {\bibfnamefont {A.}~\bibnamefont {Megrant}}, \bibinfo {author}
  {\bibfnamefont {E.}~\bibnamefont {Jeffrey}}, \bibinfo {author} {\bibfnamefont
  {T.~C.}\ \bibnamefont {White}}, \bibinfo {author} {\bibfnamefont
  {D.}~\bibnamefont {Sank}}, \bibinfo {author} {\bibfnamefont {J.~Y.}\
  \bibnamefont {Mutus}}, \bibinfo {author} {\bibfnamefont {B.}~\bibnamefont
  {Campbell}}, \bibinfo {author} {\bibfnamefont {Y.}~\bibnamefont {Chen}},
  \bibinfo {author} {\bibfnamefont {Z.}~\bibnamefont {Chen}}, \bibinfo {author}
  {\bibfnamefont {B.}~\bibnamefont {Chiaro}}, \bibinfo {author} {\bibfnamefont
  {A.}~\bibnamefont {Dunsworth}}, \bibinfo {author} {\bibfnamefont {I.-C.}\
  \bibnamefont {Hoi}}, \bibinfo {author} {\bibfnamefont {C.}~\bibnamefont
  {Neill}}, \bibinfo {author} {\bibfnamefont {P.~J.~J.}\ \bibnamefont
  {O'Malley}}, \bibinfo {author} {\bibfnamefont {C.}~\bibnamefont {Quintana}},
  \bibinfo {author} {\bibfnamefont {P.}~\bibnamefont {Roushan}}, \bibinfo
  {author} {\bibfnamefont {A.}~\bibnamefont {Vainsencher}}, \bibinfo {author}
  {\bibfnamefont {J.}~\bibnamefont {Wenner}}, \bibinfo {author} {\bibfnamefont
  {A.~N.}\ \bibnamefont {Cleland}}, \ and\ \bibinfo {author} {\bibfnamefont
  {J.~M.}\ \bibnamefont {Martinis}},\ }\href {\doibase 10.1038/nature14270}
  {\bibfield  {journal} {\bibinfo  {journal} {Nature}\ }\textbf {\bibinfo
  {volume} {519}},\ \bibinfo {pages} {66} (\bibinfo {year} {2015})}\BibitemShut
  {NoStop}%
\bibitem [{\citenamefont {Hofheinz}\ \emph {et~al.}(2009)\citenamefont
  {Hofheinz}, \citenamefont {Wang}, \citenamefont {Ansmann}, \citenamefont
  {Bialczak}, \citenamefont {Lucero}, \citenamefont {Neeley}, \citenamefont
  {O'Connell}, \citenamefont {Sank}, \citenamefont {Wenner}, \citenamefont
  {Martinis},\ and\ \citenamefont {Cleland}}]{bib:Hofheinz09}%
  \BibitemOpen
  \bibfield  {author} {\bibinfo {author} {\bibfnamefont {M.}~\bibnamefont
  {Hofheinz}}, \bibinfo {author} {\bibfnamefont {H.}~\bibnamefont {Wang}},
  \bibinfo {author} {\bibfnamefont {M.}~\bibnamefont {Ansmann}}, \bibinfo
  {author} {\bibfnamefont {R.~C.}\ \bibnamefont {Bialczak}}, \bibinfo {author}
  {\bibfnamefont {E.}~\bibnamefont {Lucero}}, \bibinfo {author} {\bibfnamefont
  {M.}~\bibnamefont {Neeley}}, \bibinfo {author} {\bibfnamefont {A.~D.}\
  \bibnamefont {O'Connell}}, \bibinfo {author} {\bibfnamefont {D.}~\bibnamefont
  {Sank}}, \bibinfo {author} {\bibfnamefont {J.}~\bibnamefont {Wenner}},
  \bibinfo {author} {\bibfnamefont {J.~M.}\ \bibnamefont {Martinis}}, \ and\
  \bibinfo {author} {\bibfnamefont {A.~N.}\ \bibnamefont {Cleland}},\ }\href
  {\doibase 10.1038/nature08005} {\bibfield  {journal} {\bibinfo  {journal}
  {Nature}\ }\textbf {\bibinfo {volume} {459}},\ \bibinfo {pages} {546}
  (\bibinfo {year} {2009})}\BibitemShut {NoStop}%
\bibitem [{\citenamefont {Flurin}\ \emph {et~al.}(2012)\citenamefont {Flurin},
  \citenamefont {Roch}, \citenamefont {Mallet}, \citenamefont {Devoret},\ and\
  \citenamefont {Huard}}]{bib:Flurin12}%
  \BibitemOpen
  \bibfield  {author} {\bibinfo {author} {\bibfnamefont {E.}~\bibnamefont
  {Flurin}}, \bibinfo {author} {\bibfnamefont {N.}~\bibnamefont {Roch}},
  \bibinfo {author} {\bibfnamefont {F.}~\bibnamefont {Mallet}}, \bibinfo
  {author} {\bibfnamefont {M.~H.}\ \bibnamefont {Devoret}}, \ and\ \bibinfo
  {author} {\bibfnamefont {B.}~\bibnamefont {Huard}},\ }\href {\doibase 10.1103/PhysRevLett.109.183901} {\bibfield  {journal} {\bibinfo  {journal}
  {Phys. Rev. Lett.}\ }\textbf {\bibinfo {volume} {109}},\ \bibinfo {pages}
  {183901} (\bibinfo {year} {2012})}\BibitemShut {NoStop}%
\bibitem [{\citenamefont {Lang}\ \emph {et~al.}(2013)\citenamefont {Lang},
  \citenamefont {Eichler}, \citenamefont {Steffen}, \citenamefont {Fink},
  \citenamefont {Woolley}, \citenamefont {Blais},\ and\ \citenamefont
  {Wallraff}}]{bib:Lang13}%
  \BibitemOpen
  \bibfield  {author} {\bibinfo {author} {\bibfnamefont {C.}~\bibnamefont
  {Lang}}, \bibinfo {author} {\bibfnamefont {C.}~\bibnamefont {Eichler}},
  \bibinfo {author} {\bibfnamefont {L.}~\bibnamefont {Steffen}}, \bibinfo
  {author} {\bibfnamefont {J.~M.}\ \bibnamefont {Fink}}, \bibinfo {author}
  {\bibfnamefont {M.~J.}\ \bibnamefont {Woolley}}, \bibinfo {author}
  {\bibfnamefont {A.}~\bibnamefont {Blais}}, \ and\ \bibinfo {author}
  {\bibfnamefont {A.}~\bibnamefont {Wallraff}},\ }\href {\doibase 10.1038/nphys2612} {\bibfield  {journal} {\bibinfo  {journal} {Nat. Phys.}\
  }\textbf {\bibinfo {volume} {9}},\ \bibinfo {pages} {345} (\bibinfo {year}
  {2013})}\BibitemShut {NoStop}%
\bibitem [{\citenamefont {Leghtas}\ \emph {et~al.}(2015)\citenamefont
  {Leghtas}, \citenamefont {Touzard}, \citenamefont {Pop}, \citenamefont {Kou},
  \citenamefont {Vlastakis}, \citenamefont {Petrenko}, \citenamefont {Sliwa},
  \citenamefont {Narla}, \citenamefont {Shankar}, \citenamefont {Hatridge},
  \citenamefont {Reagor}, \citenamefont {Frunzio}, \citenamefont {Schoelkopf},
  \citenamefont {Mirrahimi},\ and\ \citenamefont {Devoret}}]{bib:Leghtas15}%
  \BibitemOpen
  \bibfield  {author} {\bibinfo {author} {\bibfnamefont {Z.}~\bibnamefont
  {Leghtas}}, \bibinfo {author} {\bibfnamefont {S.}~\bibnamefont {Touzard}},
  \bibinfo {author} {\bibfnamefont {I.~M.}\ \bibnamefont {Pop}}, \bibinfo
  {author} {\bibfnamefont {A.}~\bibnamefont {Kou}}, \bibinfo {author}
  {\bibfnamefont {B.}~\bibnamefont {Vlastakis}}, \bibinfo {author}
  {\bibfnamefont {A.}~\bibnamefont {Petrenko}}, \bibinfo {author}
  {\bibfnamefont {K.~M.}\ \bibnamefont {Sliwa}}, \bibinfo {author}
  {\bibfnamefont {A.}~\bibnamefont {Narla}}, \bibinfo {author} {\bibfnamefont
  {S.}~\bibnamefont {Shankar}}, \bibinfo {author} {\bibfnamefont {M.~J.}\
  \bibnamefont {Hatridge}}, \bibinfo {author} {\bibfnamefont {M.}~\bibnamefont
  {Reagor}}, \bibinfo {author} {\bibfnamefont {L.}~\bibnamefont {Frunzio}},
  \bibinfo {author} {\bibfnamefont {R.~J.}\ \bibnamefont {Schoelkopf}},
  \bibinfo {author} {\bibfnamefont {M.}~\bibnamefont {Mirrahimi}}, \ and\
  \bibinfo {author} {\bibfnamefont {M.~H.}\ \bibnamefont {Devoret}},\ }\href
  {\doibase 10.1126/science.aaa2085} {\bibfield  {journal} {\bibinfo  {journal}
  {Science}\ }\textbf {\bibinfo {volume} {347}},\ \bibinfo {pages} {853}
  (\bibinfo {year} {2015})}\BibitemShut {NoStop}%
\bibitem [{Note1()}]{Note1}%
  \BibitemOpen
  \bibinfo {note} {To be more precise, we may say that surface plasmon
  polaritons, i.e., quanta of the hybridized modes of the surface charge
  density waves on the electrodes and the electromagnetic waves in the vacuum,
  are manipulated in the circuits.}\BibitemShut {Stop}%
\bibitem [{\citenamefont {Zhu}\ \emph {et~al.}(2011)\citenamefont {Zhu},
  \citenamefont {Saito}, \citenamefont {Kemp}, \citenamefont {Kakuyanagi},
  \citenamefont {Karimoto}, \citenamefont {Nakano}, \citenamefont {Munro},
  \citenamefont {Tokura}, \citenamefont {Everitt}, \citenamefont {Nemoto},
  \citenamefont {Kasu}, \citenamefont {Mizuochi},\ and\ \citenamefont
  {Semba}}]{bib:Zhu11}%
  \BibitemOpen
  \bibfield  {author} {\bibinfo {author} {\bibfnamefont {X.}~\bibnamefont
  {Zhu}}, \bibinfo {author} {\bibfnamefont {S.}~\bibnamefont {Saito}}, \bibinfo
  {author} {\bibfnamefont {A.}~\bibnamefont {Kemp}}, \bibinfo {author}
  {\bibfnamefont {K.}~\bibnamefont {Kakuyanagi}}, \bibinfo {author}
  {\bibfnamefont {S.}~\bibnamefont {Karimoto}}, \bibinfo {author}
  {\bibfnamefont {H.}~\bibnamefont {Nakano}}, \bibinfo {author} {\bibfnamefont
  {W.~J.}\ \bibnamefont {Munro}}, \bibinfo {author} {\bibfnamefont
  {Y.}~\bibnamefont {Tokura}}, \bibinfo {author} {\bibfnamefont {M.~S.}\
  \bibnamefont {Everitt}}, \bibinfo {author} {\bibfnamefont {K.}~\bibnamefont
  {Nemoto}}, \bibinfo {author} {\bibfnamefont {M.}~\bibnamefont {Kasu}},
  \bibinfo {author} {\bibfnamefont {N.}~\bibnamefont {Mizuochi}}, \ and\
  \bibinfo {author} {\bibfnamefont {K.}~\bibnamefont {Semba}},\ }\href
  {\doibase 10.1038/nature10462} {\bibfield  {journal} {\bibinfo  {journal}
  {Nature}\ }\textbf {\bibinfo {volume} {478}},\ \bibinfo {pages} {221}
  (\bibinfo {year} {2011})}\BibitemShut {NoStop}%
\bibitem [{\citenamefont {Kubo}\ \emph {et~al.}(2011)\citenamefont {Kubo},
  \citenamefont {Grezes}, \citenamefont {Dewes}, \citenamefont {Umeda},
  \citenamefont {Isoya}, \citenamefont {Sumiya}, \citenamefont {Morishita},
  \citenamefont {Abe}, \citenamefont {Onoda}, \citenamefont {Ohshima},
  \citenamefont {Jacques}, \citenamefont {Dr\'eau}, \citenamefont {Roch},
  \citenamefont {Diniz}, \citenamefont {Auffeves}, \citenamefont {Vion},
  \citenamefont {Esteve},\ and\ \citenamefont {Bertet}}]{bib:Kubo11}%
  \BibitemOpen
  \bibfield  {author} {\bibinfo {author} {\bibfnamefont {Y.}~\bibnamefont
  {Kubo}}, \bibinfo {author} {\bibfnamefont {C.}~\bibnamefont {Grezes}},
  \bibinfo {author} {\bibfnamefont {A.}~\bibnamefont {Dewes}}, \bibinfo
  {author} {\bibfnamefont {T.}~\bibnamefont {Umeda}}, \bibinfo {author}
  {\bibfnamefont {J.}~\bibnamefont {Isoya}}, \bibinfo {author} {\bibfnamefont
  {H.}~\bibnamefont {Sumiya}}, \bibinfo {author} {\bibfnamefont
  {N.}~\bibnamefont {Morishita}}, \bibinfo {author} {\bibfnamefont
  {H.}~\bibnamefont {Abe}}, \bibinfo {author} {\bibfnamefont {S.}~\bibnamefont
  {Onoda}}, \bibinfo {author} {\bibfnamefont {T.}~\bibnamefont {Ohshima}},
  \bibinfo {author} {\bibfnamefont {V.}~\bibnamefont {Jacques}}, \bibinfo
  {author} {\bibfnamefont {A.}~\bibnamefont {Dr\'eau}}, \bibinfo {author}
  {\bibfnamefont {J.-F.}\ \bibnamefont {Roch}}, \bibinfo {author}
  {\bibfnamefont {I.}~\bibnamefont {Diniz}}, \bibinfo {author} {\bibfnamefont
  {A.}~\bibnamefont {Auffeves}}, \bibinfo {author} {\bibfnamefont
  {D.}~\bibnamefont {Vion}}, \bibinfo {author} {\bibfnamefont {D.}~\bibnamefont
  {Esteve}}, \ and\ \bibinfo {author} {\bibfnamefont {P.}~\bibnamefont
  {Bertet}},\ }\href@noop {} {\bibfield  {journal} {\bibinfo  {journal} {Phys.
  Rev. Lett.}\ }\textbf {\bibinfo {volume} {107}},\ \bibinfo {pages} {220501}
  (\bibinfo {year} {2011})}\BibitemShut {NoStop}%
\bibitem [{\citenamefont {O'Connell}\ \emph {et~al.}(2010)\citenamefont
  {O'Connell}, \citenamefont {Hofheinz}, \citenamefont {Ansmann}, \citenamefont
  {Bialczak}, \citenamefont {Lenander}, \citenamefont {Lucero}, \citenamefont
  {Neeley}, \citenamefont {Sank}, \citenamefont {Wang}, \citenamefont {Weides},
  \citenamefont {Wenner}, \citenamefont {Martinis},\ and\ \citenamefont
  {Cleland}}]{bib:Connell10}%
  \BibitemOpen
  \bibfield  {author} {\bibinfo {author} {\bibfnamefont {A.~D.}\ \bibnamefont
  {O'Connell}}, \bibinfo {author} {\bibfnamefont {M.}~\bibnamefont {Hofheinz}},
  \bibinfo {author} {\bibfnamefont {M.}~\bibnamefont {Ansmann}}, \bibinfo
  {author} {\bibfnamefont {R.~C.}\ \bibnamefont {Bialczak}}, \bibinfo {author}
  {\bibfnamefont {M.}~\bibnamefont {Lenander}}, \bibinfo {author}
  {\bibfnamefont {E.}~\bibnamefont {Lucero}}, \bibinfo {author} {\bibfnamefont
  {M.}~\bibnamefont {Neeley}}, \bibinfo {author} {\bibfnamefont
  {D.}~\bibnamefont {Sank}}, \bibinfo {author} {\bibfnamefont {H.}~\bibnamefont
  {Wang}}, \bibinfo {author} {\bibfnamefont {M.}~\bibnamefont {Weides}},
  \bibinfo {author} {\bibfnamefont {J.}~\bibnamefont {Wenner}}, \bibinfo
  {author} {\bibfnamefont {J.~M.}\ \bibnamefont {Martinis}}, \ and\ \bibinfo
  {author} {\bibfnamefont {A.~N.}\ \bibnamefont {Cleland}},\ }\href {\doibase 10.1038/nature08967} {\bibfield  {journal} {\bibinfo  {journal} {Nature}\
  }\textbf {\bibinfo {volume} {464}},\ \bibinfo {pages} {697} (\bibinfo {year}
  {2010})}\BibitemShut {NoStop}%
\bibitem [{\citenamefont {Lecocq}\ \emph {et~al.}(2015)\citenamefont {Lecocq},
  \citenamefont {Teufel}, \citenamefont {Aumentado},\ and\ \citenamefont
  {Simmonds}}]{bib:Lecocq15}%
  \BibitemOpen
  \bibfield  {author} {\bibinfo {author} {\bibfnamefont {F.}~\bibnamefont
  {Lecocq}}, \bibinfo {author} {\bibfnamefont {J.~D.}\ \bibnamefont {Teufel}},
  \bibinfo {author} {\bibfnamefont {J.}~\bibnamefont {Aumentado}}, \ and\
  \bibinfo {author} {\bibfnamefont {R.~W.}\ \bibnamefont {Simmonds}},\ }\href
  {\doibase 10.1038/nphys3365} {\bibfield  {journal} {\bibinfo  {journal} {Nat.
  Phys.}\ }\textbf {\bibinfo {volume} {11}},\ \bibinfo {pages} {635} (\bibinfo
  {year} {2015})}\BibitemShut {NoStop}%
\bibitem [{\citenamefont {Pirkkalainen}\ \emph {et~al.}(2015)\citenamefont
  {Pirkkalainen}, \citenamefont {Cho}, \citenamefont {Li}, \citenamefont
  {Paraoanu}, \citenamefont {Hakonen},\ and\ \citenamefont
  {Sillanpaa}}]{bib:Pirkkalainen15}%
  \BibitemOpen
  \bibfield  {author} {\bibinfo {author} {\bibfnamefont {J.-M.}\ \bibnamefont
  {Pirkkalainen}}, \bibinfo {author} {\bibfnamefont {S.~U.}\ \bibnamefont
  {Cho}}, \bibinfo {author} {\bibfnamefont {J.}~\bibnamefont {Li}}, \bibinfo
  {author} {\bibfnamefont {G.~S.}\ \bibnamefont {Paraoanu}}, \bibinfo {author}
  {\bibfnamefont {P.~J.}\ \bibnamefont {Hakonen}}, \ and\ \bibinfo {author}
  {\bibfnamefont {M.~A.}\ \bibnamefont {Sillanpaa}},\ }\href {\doibase 10.1038/nature11821} {\bibfield  {journal} {\bibinfo  {journal} {Nature}\
  }\textbf {\bibinfo {volume} {494}},\ \bibinfo {pages} {211} (\bibinfo {year}
  {2015})}\BibitemShut {NoStop}%
\bibitem [{\citenamefont {Gustafsson}\ \emph {et~al.}(2014)\citenamefont
  {Gustafsson}, \citenamefont {Aref}, \citenamefont {Kockum}, \citenamefont
  {Ekstrom}, \citenamefont {Johansson},\ and\ \citenamefont
  {Delsing}}]{bib:Gustafsson14}%
  \BibitemOpen
  \bibfield  {author} {\bibinfo {author} {\bibfnamefont {M.~V.}\ \bibnamefont
  {Gustafsson}}, \bibinfo {author} {\bibfnamefont {T.}~\bibnamefont {Aref}},
  \bibinfo {author} {\bibfnamefont {A.~F.}\ \bibnamefont {Kockum}}, \bibinfo
  {author} {\bibfnamefont {M.~K.}\ \bibnamefont {Ekstrom}}, \bibinfo {author}
  {\bibfnamefont {G.}~\bibnamefont {Johansson}}, \ and\ \bibinfo {author}
  {\bibfnamefont {P.}~\bibnamefont {Delsing}},\ }\href {\doibase 10.1126/science.1257219} {\bibfield  {journal} {\bibinfo  {journal}
  {Science}\ }\textbf {\bibinfo {volume} {346}},\ \bibinfo {pages} {207}
  (\bibinfo {year} {2014})}\BibitemShut {NoStop}%
\bibitem [{\citenamefont {Tabuchi}\ \emph {et~al.}(2014)\citenamefont
  {Tabuchi}, \citenamefont {Ishino}, \citenamefont {Ishikawa}, \citenamefont
  {Yamazaki}, \citenamefont {Usami},\ and\ \citenamefont
  {Nakamura}}]{bib:Tabuchi14}%
  \BibitemOpen
  \bibfield  {author} {\bibinfo {author} {\bibfnamefont {Y.}~\bibnamefont
  {Tabuchi}}, \bibinfo {author} {\bibfnamefont {S.}~\bibnamefont {Ishino}},
  \bibinfo {author} {\bibfnamefont {T.}~\bibnamefont {Ishikawa}}, \bibinfo
  {author} {\bibfnamefont {R.}~\bibnamefont {Yamazaki}}, \bibinfo {author}
  {\bibfnamefont {K.}~\bibnamefont {Usami}}, \ and\ \bibinfo {author}
  {\bibfnamefont {Y.}~\bibnamefont {Nakamura}},\ }\href {\doibase 10.1103/PhysRevLett.113.083603} {\bibfield  {journal} {\bibinfo  {journal}
  {Phys. Rev. Lett.}\ }\textbf {\bibinfo {volume} {113}},\ \bibinfo {pages}
  {083603} (\bibinfo {year} {2014})}\BibitemShut {NoStop}%
\bibitem [{\citenamefont {Tabuchi}\ \emph {et~al.}(2015)\citenamefont
  {Tabuchi}, \citenamefont {Ishino}, \citenamefont {Noguchi}, \citenamefont
  {Ishikawa}, \citenamefont {Yamazaki}, \citenamefont {Usami},\ and\
  \citenamefont {Nakamura}}]{bib:Tabuchi15}%
  \BibitemOpen
  \bibfield  {author} {\bibinfo {author} {\bibfnamefont {Y.}~\bibnamefont
  {Tabuchi}}, \bibinfo {author} {\bibfnamefont {S.}~\bibnamefont {Ishino}},
  \bibinfo {author} {\bibfnamefont {A.}~\bibnamefont {Noguchi}}, \bibinfo
  {author} {\bibfnamefont {T.}~\bibnamefont {Ishikawa}}, \bibinfo {author}
  {\bibfnamefont {R.}~\bibnamefont {Yamazaki}}, \bibinfo {author}
  {\bibfnamefont {K.}~\bibnamefont {Usami}}, \ and\ \bibinfo {author}
  {\bibfnamefont {Y.}~\bibnamefont {Nakamura}},\ }\href {\doibase 10.1126/science.aaa3693} {\bibfield  {journal} {\bibinfo  {journal}
  {Science}\ }\textbf {\bibinfo {volume} {349}},\ \bibinfo {pages} {405}
  (\bibinfo {year} {2015})}\BibitemShut {NoStop}%
\bibitem [{\citenamefont {Holstein}\ and\ \citenamefont
  {Primakoff}(1940)}]{bib:Holstein40}%
  \BibitemOpen
  \bibfield  {author} {\bibinfo {author} {\bibfnamefont {T.}~\bibnamefont
  {Holstein}}\ and\ \bibinfo {author} {\bibfnamefont {H.}~\bibnamefont
  {Primakoff}},\ }\href {\doibase 10.1103/PhysRev.58.1098} {\bibfield
  {journal} {\bibinfo  {journal} {Phys. Rev.}\ }\textbf {\bibinfo {volume}
  {58}},\ \bibinfo {pages} {1098} (\bibinfo {year} {1940})}\BibitemShut
  {NoStop}%
\bibitem [{\citenamefont {Walker}(1957)}]{bib:Walker57}%
  \BibitemOpen
  \bibfield  {author} {\bibinfo {author} {\bibfnamefont {L.~R.}\ \bibnamefont
  {Walker}},\ }\href {\doibase 10.1103/PhysRev.105.390} {\bibfield  {journal}
  {\bibinfo  {journal} {Phys. Rev.}\ }\textbf {\bibinfo {volume} {105}},\
  \bibinfo {pages} {390} (\bibinfo {year} {1957})}\BibitemShut {NoStop}%
\bibitem [{\citenamefont {Walker}(1958)}]{bib:Walker58}%
  \BibitemOpen
  \bibfield  {author} {\bibinfo {author} {\bibfnamefont {L.~R.}\ \bibnamefont
  {Walker}},\ }\href {\doibase http://dx.doi.org/10.1063/1.1723117} {\bibfield
  {journal} {\bibinfo  {journal} {J. Appl. Phys.}\ }\textbf {\bibinfo {volume}
  {29}},\ \bibinfo {pages} {318} (\bibinfo {year} {1958})}\BibitemShut
  {NoStop}%
\bibitem [{\citenamefont {Fletcher}\ and\ \citenamefont
  {Bell}(1959)}]{bib:Fletcher59}%
  \BibitemOpen
  \bibfield  {author} {\bibinfo {author} {\bibfnamefont {P.~C.}\ \bibnamefont
  {Fletcher}}\ and\ \bibinfo {author} {\bibfnamefont {R.~O.}\ \bibnamefont
  {Bell}},\ }\href {\doibase http://dx.doi.org/10.1063/1.1735216} {\bibfield
  {journal} {\bibinfo  {journal} {J. Appl. Phys.}\ }\textbf {\bibinfo {volume}
  {30}},\ \bibinfo {pages} {687} (\bibinfo {year} {1959})}\BibitemShut
  {NoStop}%
\bibitem [{\citenamefont {Sparks}(1964)}]{bib:Sparks64}%
  \BibitemOpen
  \bibfield  {author} {\bibinfo {author} {\bibfnamefont {M.}~\bibnamefont
  {Sparks}},\ }\href@noop {} { {\bibinfo {title} {Ferromagnetic-relaxation
  theory}}}\ (\bibinfo  {publisher} {McGraw-Hill},\ \bibinfo {year}
  {1964})\BibitemShut {NoStop}%
\bibitem [{\citenamefont {Gurevich}\ and\ \citenamefont
  {Melkov}(1996)}]{bib:Gurevich96}%
  \BibitemOpen
  \bibfield  {author} {\bibinfo {author} {\bibfnamefont {A.~G.}\ \bibnamefont
  {Gurevich}}\ and\ \bibinfo {author} {\bibfnamefont {G.~A.}\ \bibnamefont
  {Melkov}},\ }\href@noop {} { {\bibinfo {title} {Magnetization
  oscillations and waves}}}\ (\bibinfo  {publisher} {CRC press},\ \bibinfo
  {year} {1996})\BibitemShut {NoStop}%
\bibitem [{\citenamefont {Kasuya}\ and\ \citenamefont
  {LeCraw}(1961)}]{bib:Kasuya61}%
  \BibitemOpen
  \bibfield  {author} {\bibinfo {author} {\bibfnamefont {T.}~\bibnamefont
  {Kasuya}}\ and\ \bibinfo {author} {\bibfnamefont {R.~C.}\ \bibnamefont
  {LeCraw}},\ }\href {\doibase 10.1103/PhysRevLett.6.223} {\bibfield  {journal}
  {\bibinfo  {journal} {Phys. Rev. Lett.}\ }\textbf {\bibinfo {volume} {6}},\
  \bibinfo {pages} {223} (\bibinfo {year} {1961})}\BibitemShut {NoStop}%
\bibitem [{\citenamefont {Teale}\ and\ \citenamefont
  {Tweedale}(1962)}]{bib:Teale62}%
  \BibitemOpen
  \bibfield  {author} {\bibinfo {author} {\bibfnamefont {R.}~\bibnamefont
  {Teale}}\ and\ \bibinfo {author} {\bibfnamefont {K.}~\bibnamefont
  {Tweedale}},\ }\href {\doibase 10.1016/0031-9163(62)91392-6} {\bibfield
  {journal} {\bibinfo  {journal} {{Phys. Lett.}}\ }\textbf {\bibinfo {volume}
  {1}},\ \bibinfo {pages} {298} (\bibinfo {year} {1962})}\BibitemShut {NoStop}%
\bibitem [{\citenamefont {Van~Vleck}(1964)}]{bib:Vleck64}%
  \BibitemOpen
  \bibfield  {author} {\bibinfo {author} {\bibfnamefont {J.~H.}\ \bibnamefont
  {Van~Vleck}},\ }\href {\doibase 10.1063/1.1713520} {\bibfield  {journal}
  {\bibinfo  {journal} {{J. Appl. Phys.}}\ }\textbf {\bibinfo {volume} {35}},\
  \bibinfo {pages} {882} (\bibinfo {year} {1964})}\BibitemShut {NoStop}%
\bibitem [{\citenamefont {Sparks}\ \emph {et~al.}(1961)\citenamefont {Sparks},
  \citenamefont {Loudon},\ and\ \citenamefont {Kittel}}]{bib:Sparks61}%
  \BibitemOpen
  \bibfield  {author} {\bibinfo {author} {\bibfnamefont {M.}~\bibnamefont
  {Sparks}}, \bibinfo {author} {\bibfnamefont {R.}~\bibnamefont {Loudon}}, \
  and\ \bibinfo {author} {\bibfnamefont {C.}~\bibnamefont {Kittel}},\ }\href
  {\doibase 10.1103/PhysRev.122.791} {\bibfield  {journal} {\bibinfo  {journal}
  {Phys. Rev.}\ }\textbf {\bibinfo {volume} {122}},\ \bibinfo {pages} {791}
  (\bibinfo {year} {1961})}\BibitemShut {NoStop}%
\bibitem [{\citenamefont {Huebl}\ \emph {et~al.}(2013)\citenamefont {Huebl},
  \citenamefont {Zollitsch}, \citenamefont {Lotze}, \citenamefont {Hocke},
  \citenamefont {Greifenstein}, \citenamefont {Marx}, \citenamefont {Gross},\
  and\ \citenamefont {Goennenwein}}]{bib:Huebl13}%
  \BibitemOpen
  \bibfield  {author} {\bibinfo {author} {\bibfnamefont {H.}~\bibnamefont
  {Huebl}}, \bibinfo {author} {\bibfnamefont {C.~W.}\ \bibnamefont
  {Zollitsch}}, \bibinfo {author} {\bibfnamefont {J.}~\bibnamefont {Lotze}},
  \bibinfo {author} {\bibfnamefont {F.}~\bibnamefont {Hocke}}, \bibinfo
  {author} {\bibfnamefont {M.}~\bibnamefont {Greifenstein}}, \bibinfo {author}
  {\bibfnamefont {A.}~\bibnamefont {Marx}}, \bibinfo {author} {\bibfnamefont
  {R.}~\bibnamefont {Gross}}, \ and\ \bibinfo {author} {\bibfnamefont
  {S.~T.~B.}\ \bibnamefont {Goennenwein}},\ }\href {\doibase 10.1103/PhysRevLett.111.127003} {\bibfield  {journal} {\bibinfo  {journal}
  {Phys. Rev. Lett.}\ }\textbf {\bibinfo {volume} {111}},\ \bibinfo {pages}
  {127003} (\bibinfo {year} {2013})}\BibitemShut {NoStop}%
\bibitem [{\citenamefont {Zhang}\ \emph {et~al.}(2014)\citenamefont {Zhang},
  \citenamefont {Zou}, \citenamefont {Jiang},\ and\ \citenamefont
  {Tang}}]{bib:Zhang14}%
  \BibitemOpen
  \bibfield  {author} {\bibinfo {author} {\bibfnamefont {X.}~\bibnamefont
  {Zhang}}, \bibinfo {author} {\bibfnamefont {C.-L.}\ \bibnamefont {Zou}},
  \bibinfo {author} {\bibfnamefont {L.}~\bibnamefont {Jiang}}, \ and\ \bibinfo
  {author} {\bibfnamefont {H.~X.}\ \bibnamefont {Tang}},\ }\href {\doibase 10.1103/PhysRevLett.113.156401} {\bibfield  {journal} {\bibinfo  {journal}
  {Phys. Rev. Lett.}\ }\textbf {\bibinfo {volume} {113}},\ \bibinfo {pages}
  {156401} (\bibinfo {year} {2014})}\BibitemShut {NoStop}%
\bibitem [{\citenamefont {Goryachev}\ \emph {et~al.}(2014)\citenamefont
  {Goryachev}, \citenamefont {Farr}, \citenamefont {Creedon}, \citenamefont
  {Fan}, \citenamefont {Kostylev},\ and\ \citenamefont
  {Tobar}}]{bib:Goryachev14}%
  \BibitemOpen
  \bibfield  {author} {\bibinfo {author} {\bibfnamefont {M.}~\bibnamefont
  {Goryachev}}, \bibinfo {author} {\bibfnamefont {W.~G.}\ \bibnamefont {Farr}},
  \bibinfo {author} {\bibfnamefont {D.~L.}\ \bibnamefont {Creedon}}, \bibinfo
  {author} {\bibfnamefont {Y.}~\bibnamefont {Fan}}, \bibinfo {author}
  {\bibfnamefont {M.}~\bibnamefont {Kostylev}}, \ and\ \bibinfo {author}
  {\bibfnamefont {M.~E.}\ \bibnamefont {Tobar}},\ }\href {\doibase 10.1103/PhysRevApplied.2.054002} {\bibfield  {journal} {\bibinfo  {journal}
  {Phys. Rev. Applied}\ }\textbf {\bibinfo {volume} {2}},\ \bibinfo {pages}
  {054002} (\bibinfo {year} {2014})}\BibitemShut {NoStop}%
\bibitem [{\citenamefont {Cherepanov}\ \emph {et~al.}(1993)\citenamefont
  {Cherepanov}, \citenamefont {Kolokolov},\ and\ \citenamefont
  {L'vov}}]{bib:Cherepanov93}%
  \BibitemOpen
  \bibfield  {author} {\bibinfo {author} {\bibfnamefont {V.}~\bibnamefont
  {Cherepanov}}, \bibinfo {author} {\bibfnamefont {I.}~\bibnamefont
  {Kolokolov}}, \ and\ \bibinfo {author} {\bibfnamefont {V.}~\bibnamefont
  {L'vov}},\ }\href@noop {} {\bibfield  {journal} {\bibinfo  {journal} {Physics
  Reports}\ }\textbf {\bibinfo {volume} {229}},\ \bibinfo {pages} {81}
  (\bibinfo {year} {1993})}\BibitemShut {NoStop}%
\bibitem [{\citenamefont {Demokritov}\ \emph {et~al.}(2006)\citenamefont
  {Demokritov}, \citenamefont {Demidov}, \citenamefont {Dzyapko}, \citenamefont
  {Melkov}, \citenamefont {Serga}, \citenamefont {Hillebrands},\ and\
  \citenamefont {Slavin}}]{bib:Demokritov06}%
  \BibitemOpen
  \bibfield  {author} {\bibinfo {author} {\bibfnamefont {S.~O.}\ \bibnamefont
  {Demokritov}}, \bibinfo {author} {\bibfnamefont {V.~E.}\ \bibnamefont
  {Demidov}}, \bibinfo {author} {\bibfnamefont {O.}~\bibnamefont {Dzyapko}},
  \bibinfo {author} {\bibfnamefont {G.~A.}\ \bibnamefont {Melkov}}, \bibinfo
  {author} {\bibfnamefont {A.~A.}\ \bibnamefont {Serga}}, \bibinfo {author}
  {\bibfnamefont {B.}~\bibnamefont {Hillebrands}}, \ and\ \bibinfo {author}
  {\bibfnamefont {A.~N.}\ \bibnamefont {Slavin}},\ }\href {\doibase 10.1038/nature05117} {\bibfield  {journal} {\bibinfo  {journal} {Nature}\
  }\textbf {\bibinfo {volume} {443}},\ \bibinfo {pages} {430} (\bibinfo {year}
  {2006})}\BibitemShut {NoStop}%
\bibitem [{\citenamefont {Uchida}\ \emph {et~al.}(2010)\citenamefont {Uchida},
  \citenamefont {Xiao}, \citenamefont {Adachi}, \citenamefont {Ohe},
  \citenamefont {Takahashi}, \citenamefont {Ieda}, \citenamefont {Ota},
  \citenamefont {Kajiwara}, \citenamefont {Umezawa}, \citenamefont {Kawai},
  \citenamefont {Bauer}, \citenamefont {Maekawa},\ and\ \citenamefont
  {Saitoh}}]{bib:Uchida10}%
  \BibitemOpen
  \bibfield  {author} {\bibinfo {author} {\bibfnamefont {K.}~\bibnamefont
  {Uchida}}, \bibinfo {author} {\bibfnamefont {J.}~\bibnamefont {Xiao}},
  \bibinfo {author} {\bibfnamefont {H.}~\bibnamefont {Adachi}}, \bibinfo
  {author} {\bibfnamefont {J.}~\bibnamefont {Ohe}}, \bibinfo {author}
  {\bibfnamefont {S.}~\bibnamefont {Takahashi}}, \bibinfo {author}
  {\bibfnamefont {J.}~\bibnamefont {Ieda}}, \bibinfo {author} {\bibfnamefont
  {T.}~\bibnamefont {Ota}}, \bibinfo {author} {\bibfnamefont {Y.}~\bibnamefont
  {Kajiwara}}, \bibinfo {author} {\bibfnamefont {H.}~\bibnamefont {Umezawa}},
  \bibinfo {author} {\bibfnamefont {H.}~\bibnamefont {Kawai}}, \bibinfo
  {author} {\bibfnamefont {G.~E.~W.}\ \bibnamefont {Bauer}}, \bibinfo {author}
  {\bibfnamefont {S.}~\bibnamefont {Maekawa}}, \ and\ \bibinfo {author}
  {\bibfnamefont {E.}~\bibnamefont {Saitoh}},\ }\href {\doibase 10.1038/nmat2856} {\bibfield  {journal} {\bibinfo  {journal} {Nat. Matter.}\
  }\textbf {\bibinfo {volume} {9}},\ \bibinfo {pages} {894} (\bibinfo {year}
  {2010})}\BibitemShut {NoStop}%
\bibitem [{\citenamefont {Kajiwara}\ \emph {et~al.}(2010)\citenamefont
  {Kajiwara}, \citenamefont {Harii}, \citenamefont {Takahashi}, \citenamefont
  {Ohe}, \citenamefont {Uchida}, \citenamefont {Mizuguchi}, \citenamefont
  {Umezawa}, \citenamefont {Kawai}, \citenamefont {Ando}, \citenamefont
  {Takanashi}, \citenamefont {Maekawa},\ and\ \citenamefont
  {Saitoh}}]{bib:Kajiwara10}%
  \BibitemOpen
  \bibfield  {author} {\bibinfo {author} {\bibfnamefont {Y.}~\bibnamefont
  {Kajiwara}}, \bibinfo {author} {\bibfnamefont {K.}~\bibnamefont {Harii}},
  \bibinfo {author} {\bibfnamefont {S.}~\bibnamefont {Takahashi}}, \bibinfo
  {author} {\bibfnamefont {J.}~\bibnamefont {Ohe}}, \bibinfo {author}
  {\bibfnamefont {K.}~\bibnamefont {Uchida}}, \bibinfo {author} {\bibfnamefont
  {M.}~\bibnamefont {Mizuguchi}}, \bibinfo {author} {\bibfnamefont
  {H.}~\bibnamefont {Umezawa}}, \bibinfo {author} {\bibfnamefont
  {H.}~\bibnamefont {Kawai}}, \bibinfo {author} {\bibfnamefont
  {K.}~\bibnamefont {Ando}}, \bibinfo {author} {\bibfnamefont {K.}~\bibnamefont
  {Takanashi}}, \bibinfo {author} {\bibfnamefont {S.}~\bibnamefont {Maekawa}},
  \ and\ \bibinfo {author} {\bibfnamefont {E.}~\bibnamefont {Saitoh}},\ }\href
  {\doibase 10.1038/nature08876} {\bibfield  {journal} {\bibinfo  {journal}
  {Nature}\ }\textbf {\bibinfo {volume} {464}},\ \bibinfo {pages} {262}
  (\bibinfo {year} {2010})}\BibitemShut {NoStop}%
\bibitem [{\citenamefont {O'Connell}\ \emph {et~al.}(2008)\citenamefont
  {O'Connell}, \citenamefont {Ansmann}, \citenamefont {Bialczak}, \citenamefont
  {Hofheinz}, \citenamefont {Katz}, \citenamefont {Lucero}, \citenamefont
  {McKenney}, \citenamefont {Neeley}, \citenamefont {Wang}, \citenamefont
  {Weig}, \citenamefont {Cleland},\ and\ \citenamefont
  {Martinis}}]{bib:OConnell08}%
  \BibitemOpen
  \bibfield  {author} {\bibinfo {author} {\bibfnamefont {A.~D.}\ \bibnamefont
  {O'Connell}}, \bibinfo {author} {\bibfnamefont {M.}~\bibnamefont {Ansmann}},
  \bibinfo {author} {\bibfnamefont {R.~C.}\ \bibnamefont {Bialczak}}, \bibinfo
  {author} {\bibfnamefont {M.}~\bibnamefont {Hofheinz}}, \bibinfo {author}
  {\bibfnamefont {N.}~\bibnamefont {Katz}}, \bibinfo {author} {\bibfnamefont
  {E.}~\bibnamefont {Lucero}}, \bibinfo {author} {\bibfnamefont
  {C.}~\bibnamefont {McKenney}}, \bibinfo {author} {\bibfnamefont
  {M.}~\bibnamefont {Neeley}}, \bibinfo {author} {\bibfnamefont
  {H.}~\bibnamefont {Wang}}, \bibinfo {author} {\bibfnamefont {E.~M.}\
  \bibnamefont {Weig}}, \bibinfo {author} {\bibfnamefont {A.~N.}\ \bibnamefont
  {Cleland}}, \ and\ \bibinfo {author} {\bibfnamefont {J.~M.}\ \bibnamefont
  {Martinis}},\ }\href {\doibase http://dx.doi.org/10.1063/1.2898887}
  {\bibfield  {journal} {\bibinfo  {journal} {Appl. Phys. Lett.}\ }\textbf
  {\bibinfo {volume} {92}},\ \bibinfo {eid} {112903} (\bibinfo {year} {2008}),\
  http://dx.doi.org/10.1063/1.2898887}\BibitemShut {NoStop}%
\bibitem [{\citenamefont {Gao}(2008)}]{bib:Jiansong08}%
  \BibitemOpen
  \bibfield  {author} {\bibinfo {author} {\bibfnamefont {J.}~\bibnamefont
  {Gao}},\ } {\bibinfo {title} {The Physics of Superconducting Microwave
  Resonators}},\ \href {http://thesis.library.caltech.edu/2530/} {\bibinfo
  {type} {{PhD} dissertation}},\ \bibinfo  {school} {California Institude of
  Technology} (\bibinfo {year} {2008})\BibitemShut {NoStop}%
\bibitem [{\citenamefont {Scholz}\ \emph {et~al.}(2010)\citenamefont {Scholz},
  \citenamefont {van Beek},\ and\ \citenamefont {Ernst}}]{bib:Scholz10}%
  \BibitemOpen
  \bibfield  {author} {\bibinfo {author} {\bibfnamefont {I.}~\bibnamefont
  {Scholz}}, \bibinfo {author} {\bibfnamefont {J.~D.}\ \bibnamefont {van
  Beek}}, \ and\ \bibinfo {author} {\bibfnamefont {M.}~\bibnamefont {Ernst}},\
  }\href {\doibase 10.1016/j.ssnmr.2010.04.003} {\bibfield  {journal} {\bibinfo
   {journal} {Solid State Nucl. Mag.}\ }\textbf {\bibinfo {volume} {37}},\
  \bibinfo {pages} {39} (\bibinfo {year} {2010})}\BibitemShut {NoStop}%
\bibitem [{\citenamefont {Imamo\ifmmode~\breve{g}\else
  \u{g}\fi{}lu}(2009)}]{bib:Imamoglu09}%
  \BibitemOpen
  \bibfield  {author} {\bibinfo {author} {\bibfnamefont {A.}~\bibnamefont
  {Imamo\ifmmode~\breve{g}\else \u{g}\fi{}lu}},\ }\href {\doibase 10.1103/PhysRevLett.102.083602} {\bibfield  {journal} {\bibinfo  {journal}
  {Phys. Rev. Lett.}\ }\textbf {\bibinfo {volume} {102}},\ \bibinfo {pages}
  {083602} (\bibinfo {year} {2009})}\BibitemShut {NoStop}%
\bibitem [{\citenamefont {Longdell}\ \emph {et~al.}(2005)\citenamefont
  {Longdell}, \citenamefont {Fraval}, \citenamefont {Sellars},\ and\
  \citenamefont {Manson}}]{bib:Jevon05}%
  \BibitemOpen
  \bibfield  {author} {\bibinfo {author} {\bibfnamefont {J.~J.}\ \bibnamefont
  {Longdell}}, \bibinfo {author} {\bibfnamefont {E.}~\bibnamefont {Fraval}},
  \bibinfo {author} {\bibfnamefont {M.~J.}\ \bibnamefont {Sellars}}, \ and\
  \bibinfo {author} {\bibfnamefont {N.~B.}\ \bibnamefont {Manson}},\ }\href
  {\doibase 10.1103/PhysRevLett.95.063601} {\bibfield  {journal} {\bibinfo
  {journal} {Phys. Rev. Lett.}\ }\textbf {\bibinfo {volume} {95}},\ \bibinfo
  {pages} {063601} (\bibinfo {year} {2005})}\BibitemShut {NoStop}%
\bibitem [{\citenamefont {Hammerer}\ \emph {et~al.}(2010)\citenamefont
  {Hammerer}, \citenamefont {S\o{}rensen},\ and\ \citenamefont
  {Polzik}}]{bib:Hammerer10}%
  \BibitemOpen
  \bibfield  {author} {\bibinfo {author} {\bibfnamefont {K.}~\bibnamefont
  {Hammerer}}, \bibinfo {author} {\bibfnamefont {A.~S.}\ \bibnamefont
  {S\o{}rensen}}, \ and\ \bibinfo {author} {\bibfnamefont {E.~S.}\ \bibnamefont
  {Polzik}},\ }\href {\doibase 10.1103/RevModPhys.82.1041} {\bibfield
  {journal} {\bibinfo  {journal} {Rev. Mod. Phys.}\ }\textbf {\bibinfo {volume}
  {82}},\ \bibinfo {pages} {1041} (\bibinfo {year} {2010})}\BibitemShut
  {NoStop}%
\end{thebibliography}%

\end{document}